\documentclass[10pt,letterpaper]{article}
\usepackage{geometry}
\usepackage{amsmath,slashed,bbm}%,mathrsfs,psfrag,relsize
\usepackage{setspace}
\usepackage{amssymb}
\csname@addtoreset\endcsname{equation}{section}

\makeatletter
\pdfoutput=1

\usepackage[T1]{fontenc}
\usepackage{ae}
\usepackage{aecompl}

\usepackage[english]{babel}
\usepackage[sf,bf,small]{titlesec}
\usepackage[small,sf,bf,hang]{caption}

\usepackage[multiple]{footmisc}
\long\def\symbolfootnote[#1]#2{\begingroup%
\def\thefootnote{\fnsymbol{footnote}}\footnote[#1]{#2}\endgroup}

\usepackage{titletoc}
\dottedcontents{section}[6mm]{\smallskip }{6mm}{0pc} 
\dottedcontents{subsection}[8mm]{\itshape\small }{6mm}{0pc}
\def\tableofcontents{\subsection*{\contentsname}\vspace{-2mm}\@starttoc{toc}}

\usepackage[nosort]{cite} 

\usepackage[parsep]{collref}

\usepackage{graphicx}   % auto-detectable
\usepackage{epstopdf}
\usepackage{feynmp}  
\renewcommand{\bar}[1]{\overline{#1}}
\def \bea  {\begin{eqnarray}}
\def \eea  {\end{eqnarray}}

\newcommand{\ket}[1]{|{#1}\rangle}

\newcommand{\nn}{\nonumber}

\newcommand{\ns} \normalsize

\AtBeginDocument{
  
}

\makeatother

\begin{document}
%\selectlanguage{english}
\begin{flushright}
MIFPA-12-17
\bigskip\bigskip\bigskip
\par\end{flushright}

\begin{center}
\textsf{\textbf{\Large Near BMN dynamics of the $AdS_3\times S^3\times S^3\times S^1$ superstring\smallskip\smallskip }}\\
\textsf{\textbf{\Large  }}
\par\end{center}{\Large \par}

\begin{singlespace}
\begin{center}
\large Nitin Rughoonauth$^{1}$, Per Sundin$^{1}$ and  Linus Wulff$^{2}$ \bigskip \\

{\small $^{1}$}\emph{\small{} Astrophysics, Cosmology \& Gravity
Center }\\
\emph{\small and Department of Applied Mathematics, }\\
\emph{\small University of Cape Town,}\\
\emph{\small Private Bag, Rondebosch, 7700, South Africa}{\small }\\
\emph{\small nitincr@gmail.com, nidnus.rep@gmail.com}\vspace{0.5cm}

{\small $^{2}$}\emph{\small{} George P. \& Cynthia Woods Mitchell Institute for Fundamental Physics and Astronomy,}\\
\emph{\small Texas A\&M University, College Station, }\\
\emph{\small TX 77843, USA}\\
\emph{\small linus@physics.tamu.edu }{\small \bigskip }\emph{\small }\\
\par\end{center}{\small \par}
\end{singlespace}

\subsection*{\hspace{9mm}Abstract}
\begin{quote}
We investigate the type IIA $AdS_3\times S^3 \times M_4$ superstring with $M_4=S^3\times S^1$ or $M_4=T^4$. String theory in this background is interesting because of $AdS_3 / CFT_2$ and its newly discovered integrable structures. We derive the kappa symmetry gauge-fixed Green-Schwarz string action to quadratic order in fermions and quartic order in fields utilizing a near BMN expansion. As a first consistency check of our results we show that the two point functions are one-loop finite in dimensional regularization. We then perform a Hamiltonian analysis where we compare the energy of string states with the predictions of a set of conjectured Bethe equations. While we find perfect agreement for single rank one sectors, we find that the product $SU(2)\times SU(2)$ sector does not match unless the Bethe equations decouple completely. We then calculate $2\rightarrow 2$ bosonic tree-level scattering processes on the string worldsheet and show that the two-dimensional S-matrix is reflectionless. This might be important due to the presence of massless worldsheet excitations which are generally not described by the Bethe equations.

\bigskip  \thispagestyle{empty} 
\end{quote}
%%
%% This file contains descriptions of feynman diagrams for use with the package feynMP
%% After running the main file, run "mpost diagrams" once to generate these
%%
%%
%% These are for propagators paper with Per. 
%% The version in this folder edited 18 May 2011 to add unlabled ones for introduction. 
%%

\newsavebox{\feynmanrules}
\sbox{\feynmanrules}{
\begin{fmffile}{diagrams} % I can't seem to make this work using any path but the same one as the document

%%%%%%%%%%%%%%%%
%%  SETTINGS

\fmfset{thin}{0.6pt}  % was 0.7 until v24
%\fmfset{wiggly_len}{5mm}
\fmfset{dash_len}{4pt}
\fmfset{dot_size}{1thick}
\fmfset{arrow_len}{6pt} % you can't use em here, mpost doesn't know what it will be.
%\fmfset{curly_len}{2.5mm}
%\setlength{\unitlength}{1em} % default is =1pt, maybe that's sensible. 72pt = 1in

%%%%%%%%%%%%%%%%%%%
%% HEAVY-MODE DECAY PROCESSES

\begin{fmfgraph*}(100,36)
\fmfkeep{decay1}
\fmfleft{v1}
\fmfright{o1,o2}
\fmfdot{v2}
\fmf{dashes_arrow,label=\small{$y_1$}}{v1,v2}
\fmf{fermion}{v2,o2}
\fmf{fermion}{v2,o1}
\fmflabel{\small{$\chi_\pm^{(3)}$}}{o1}
\fmflabel{\small{$\chi_\pm^{(2)}$}}{o2}
\end{fmfgraph*}

\begin{fmfgraph*}(100,36)
\fmfkeep{decay2}
\fmfleft{v1}
\fmfright{o1,o2}
\fmfdot{v2}
\fmf{fermion,label=\small{$\chi_\pm^{(1)}$}}{v1,v2}
\fmf{dashes_arrow}{v2,o2}
\fmf{fermion}{v2,o1}
\fmflabel{\small{$\chi_\pm^{(3)}$}}{o1}
\fmflabel{\small{$y_2$}}{o2}
\end{fmfgraph*}

\begin{fmfgraph*}(100,36)
\fmfkeep{decay3}
\fmfleft{v1}
\fmfright{o1,o2}
\fmfdot{v2}
\fmf{fermion,label=\small{$\chi_\pm^{(1)}$}}{v1,v2}
\fmf{dashes_arrow}{v2,o2}
\fmf{fermion}{v2,o1}
\fmflabel{\small{$\chi_\pm^{(2)}$}}{o1}
\fmflabel{\small{$y_3$}}{o2}
\end{fmfgraph*}

%%%%%%%%%%%%%%%%%%%%

%% TREE LEVEL SCATTERING

\begin{fmfgraph*}(50,25)
\fmfkeep{S-t-chan}
\fmfbottom{i1,d1,o1}
\fmftop{i2,d2,o2}
\fmf{fermion}{i1,v1,o1}
\fmf{fermion}{i2,v2,o2}
\fmf{plain,tension=0}{v1,v2}
\fmflabel{$(t)$}{v1}
\fmflabel{$2$}{i1}
\fmflabel{$1$}{i2}
\fmflabel{$4$}{o1}
\fmflabel{$3$}{o2}
\end{fmfgraph*}

\begin{fmfgraph*}(50,25)
\fmfkeep{S-u-chan}
\fmfbottom{i1,d1,o1}
\fmftop{i2,d2,o2}
\fmf{fermion,tension=2}{i1,v1}
\fmf{phantom,tension=1.5}{v1,o1}
\fmf{plain}{v1,v3}
\fmf{fermion}{v3,o2}
\fmf{fermion,tension=2}{i2,v2}
\fmf{phantom,tension=1.5}{v2,o2}
\fmf{plain}{v2,v3}
\fmf{fermion}{v3,o1}
\fmf{plain,tension=-0.5}{v1,v2}
\fmflabel{$(u)$}{v1}
\fmflabel{$2$}{i1}
\fmflabel{$1$}{i2}
\fmflabel{$4$}{o1}
\fmflabel{$3$}{o2}
\end{fmfgraph*}

\begin{fmfgraph*}(50,25)
\fmfkeep{T-t-chan}
\fmfbottom{i1,d1,o1}
\fmftop{i2,d2,o2}
\fmf{fermion}{o1,v1,i1}
\fmf{fermion}{i2,v2,o2}
\fmf{plain,tension=0}{v1,v2}
\fmflabel{$(t)$}{v1}
\fmflabel{$2$}{i1}
\fmflabel{$1$}{i2}
\fmflabel{$4$}{o1}
\fmflabel{$3$}{o2}
\end{fmfgraph*}

\begin{fmfgraph*}(50,25)
\fmfkeep{T-s-chan}
\fmfleft{i1,i2}
\fmfright{o1,o2}
\fmf{fermion}{i2,v1,i1}
\fmf{fermion}{o1,v2,o2}
\fmf{plain,label=$(s)$,label.dist=16}{v1,v2}
\fmflabel{$2$}{i1}
\fmflabel{$1$}{i2}
\fmflabel{$4$}{o1}
\fmflabel{$3$}{o2}
\end{fmfgraph*}

\begin{fmfgraph*}(50,25)
\fmfkeep{R-t-chan}
\fmfbottom{i1,d1,o1}
\fmftop{i2,d2,o2}
\fmf{fermion}{o1,v1,i1}
\fmf{fermion}{i2,v2,o2}
\fmf{plain,tension=0}{v1,v2}
\fmflabel{$(t)$}{v1}
\fmflabel{$2$}{i1}
\fmflabel{$1$}{i2}
\fmflabel{$3$}{o1}
\fmflabel{$4$}{o2}
\end{fmfgraph*}

\begin{fmfgraph*}(50,25)
\fmfkeep{R-s-chan}
\fmfleft{i1,i2}
\fmfright{o1,o2}
\fmf{fermion}{i2,v1,i1}
\fmf{fermion}{o1,v2,o2}
\fmf{plain,label=$(s)$,label.dist=16}{v1,v2}
\fmflabel{$2$}{i1}
\fmflabel{$1$}{i2}
\fmflabel{$3$}{o1}
\fmflabel{$4$}{o2}
\end{fmfgraph*}

\begin{fmfgraph*}(50,25)
\fmfkeep{S-c}
\fmfbottom{i1,o1}
\fmftop{i2,o2}
\fmf{fermion}{i1,v1,o1}
\fmf{fermion}{i2,v1,o2}
%\fmf{plain,label=$(s)$}{v1}
%\fmf{fermion,tension=0}{v1,v2}
\fmf{phantom,label=$(c)$}{i1,o1}
\fmflabel{$(t)$}{i2}
\fmflabel{$2$}{i1}
\fmflabel{$1$}{i2}
\fmflabel{$4$}{o1}
\fmflabel{$3$}{o2}
\end{fmfgraph*}

\begin{fmfgraph*}(50,25)
\fmfkeep{T-c}
\fmfbottom{i1,o1}
\fmftop{i2,o2}
\fmf{fermion}{i2,v1,i1}
\fmf{fermion}{o1,v1,o2}
\fmf{phantom,label=$(c)$}{i1,o1}
%\fmf{fermion,tension=0}{v1,v2}
%\fmflabel{$(t)$}{v1}
\fmflabel{$2$}{i1}
\fmflabel{$1$}{i2}
\fmflabel{$4$}{o1}
\fmflabel{$3$}{o2}
\end{fmfgraph*}

\begin{fmfgraph*}(50,25)
\fmfkeep{R-c}
\fmfbottom{i1,o1}
\fmftop{i2,o2}
\fmf{fermion}{i2,v1,i1}
\fmf{fermion}{o1,v1,o2}
\fmf{phantom,label=$(c)$}{i1,o1}
%\fmf{fermion,tension=0}{v1,v2}
%\fmflabel{$(t)$}{v1}
\fmflabel{$2$}{i1}
\fmflabel{$1$}{i2}
\fmflabel{$3$}{o1}
\fmflabel{$4$}{o2}
\end{fmfgraph*}

%%%%%%%%%%%%%%%%%%%%
%% TADPOLE GRAPHS  

 \begin{fmfgraph*}(110,62)
	\fmfkeep{tadpoleaaa}
% Note that the size is given in normal parentheses
% instead of curly brackets.
% Define external vertices from bottom to top
 \fmfleft{l}
    \fmf{photon}{l,i}
    \fmf{fermion}{f1,i,f2}
    \fmfright{f1,f2}

   \end{fmfgraph*}

\begin{fmfgraph*}(72,25)
\fmfkeep{tadpole-lightlight}
\fmfleft{in,p1}
\fmfright{out,p2}
\fmfdot{c}
\fmf{dbl_dashes,label=\small{$y_i(p)$}}{in,c}
\fmf{dbl_dashes}{c,out}
\fmf{plain_arrow,right, tension=0.8, label=\small{$y_k\text{ or }\chi^k_\pm \text{ with }\Lambda$}}{c,c}
\fmf{phantom, tension=0.2}{p1,p2}
\end{fmfgraph*}

\begin{fmfgraph*}(72,25)
\fmfkeep{tadpole-heavyheavy}
\fmfleft{in,p1}
\fmfright{out,p2}
\fmfdot{c}
\fmf{dbl_dashes,label=\small{$y_i(p)$}}{in,c}
\fmf{dbl_dashes}{c,out}
\fmf{plain_arrow,right, tension=0.8, label=\small{$y_1\text{ or }\chi^1_\pm \text{ with }\alpha\Lambda$}}{c,c}
\fmf{phantom, tension=0.2}{p1,p2}
\end{fmfgraph*}

%\begin{fmfgraph*}(72,25)
%\fmfkeep{tadpole-heavy-less}
%\fmfleft{in,p1}
%\fmfright{out,p2}
%\fmfdot{c}
%\fmf{dashes_arrow,label=\small{$ \omega_\alpha (p)$}}{in,c}
%\fmf{dashes_arrow}{c,out}
%\fmf{dbl_plain,right, tension=0.8, label=\small{ \  }}{c,c}
%\fmf{phantom, tension=0.2}{p1,p2}
%\end{fmfgraph*}

%%% y ext

\begin{fmfgraph*}(72,25)
\fmfkeep{tad-nolabel}
\fmfset{dash_len}{6pt} % this seems to be a local change
\fmfleft{in,p1}
\fmfright{out,p2}
\fmfdot{c}
\fmf{dashes_arrow,label=\small{$ y_i $}}{in,c}
\fmf{dashes_arrow}{c,out}
\fmf{plain_arrow,right, tension=0.8}{c,c}
\fmf{phantom, tension=0.2}{p1,p2}
\end{fmfgraph*}

\begin{fmfgraph*}(72,25)
\fmfkeep{tadpole-yy}
\fmfset{dash_len}{6pt} % this seems to be a local change
\fmfleft{in,p1}
\fmfright{out,p2}
\fmfdot{c}
\fmf{dbl_dashes,label=\small{$ y $}}{in,c}
\fmf{dbl_dashes}{c,out}
\fmf{plain_arrow,right, tension=0.8, label=\small{$\omega_\alpha\text{ or }\psi^a$}}{c,c}
\fmf{phantom, tension=0.2}{p1,p2}
\end{fmfgraph*}

\begin{fmfgraph*}(72,25)
\fmfkeep{tadpole-y4}
\fmfset{dash_len}{6pt} % this seems to be a local change
\fmfleft{in,p1}
\fmfright{out,p2}
\fmfdot{c}
\fmf{dashes_arrow,label=\small{$ y_4(p) $}}{in,c}
\fmf{dashes_arrow}{c,out}
\fmf{plain_arrow,right, tension=0.8, label=\small{$y_k$}}{c,c}
\fmf{phantom, tension=0.2}{p1,p2}
\end{fmfgraph*}

\begin{fmfgraph*}(72,25)
\fmfkeep{tadpole-y2bos}
\fmfset{dash_len}{6pt} % this seems to be a local change
\fmfleft{in,p1}
\fmfright{out,p2}
\fmfdot{c}
\fmf{dashes_arrow,label=\small{$ y_i(p) $}}{in,c}
\fmf{dashes_arrow}{c,out}
\fmf{plain_arrow,right, tension=0.8, label=\small{$y$}}{c,c}
\fmf{phantom, tension=0.2}{p1,p2}
\end{fmfgraph*}

\begin{fmfgraph*}(72,25)
\fmfkeep{tadpole-y2ferm}
\fmfset{dash_len}{6pt} % this seems to be a local change
\fmfleft{in,p1}
\fmfright{out,p2}
\fmfdot{c}
\fmf{dashes_arrow,label=\small{$ y_i(p) $}}{in,c}
\fmf{dashes_arrow}{c,out}
\fmf{plain_arrow,right, tension=0.8, label=\small{$\chi_\pm$}}{c,c}
\fmf{phantom, tension=0.2}{p1,p2}
\end{fmfgraph*}
%%% lollipop

\begin{fmfgraph*}(72,36)
\fmfkeep{lollipop}

\fmfstraight
\fmfleft{in,i1,i2}
\fmfright{out,o1,o2}
\fmfdot{bot,mid}

\fmf{dashes_arrow}{in,bot} %  removed ,label=\small{$\omega_\alpha (p)$}
\fmf{dashes_arrow}{bot,out}

\fmf{phantom}{i2,top}
\fmf{phantom}{top,o2}
\fmffreeze

\fmf{phantom}{i1,mid}
\fmf{phantom}{mid,o1}

\fmf{plain_arrow, tension=2}{bot,mid}

\fmf{plain_arrow,right, tension=0.8}{mid,top}
\fmf{plain,right, tension=0.8}{top,mid}
\end{fmfgraph*}

\begin{fmfgraph*}(72,36)

\fmfkeep{lollF}

\fmfstraight
\fmfleft{in,i1,i2}
\fmfright{out,o1,o2}
\fmfdot{bot,mid}

\fmf{dashes_arrow}{in,bot}
\fmf{dashes_arrow}{bot,out}

\fmf{phantom}{i2,top}
\fmf{phantom}{top,o2}
\fmffreeze

\fmf{phantom}{i1,mid}
\fmf{phantom}{mid,o1}

\fmf{plain_arrow, tension=2}{bot,mid}

\fmf{plain_arrow,right, tension=0.8}{mid,top}
\fmf{plain,right, tension=0.8,label=\small{$\chi_\pm^{(i)}$}}{top,mid}
\end{fmfgraph*}

\begin{fmfgraph*}(72,36)
\fmfkeep{lollB}

\fmfstraight
\fmfleft{in,i1,i2}
\fmfright{out,o1,o2}
\fmfdot{bot,mid}

\fmf{dashes_arrow}{in,bot} %  removed ,label=\small{$\omega_\alpha (p)$}
\fmf{dashes_arrow}{bot,out}

\fmf{phantom}{i2,top}
\fmf{phantom}{top,o2}
\fmffreeze

\fmf{phantom}{i1,mid}
\fmf{phantom}{mid,o1}

\fmf{plain_arrow, tension=2}{bot,mid}

\fmf{plain_arrow,right, tension=0.8}{mid,top}
\fmf{plain,right, tension=0.8,label=\small{$y_{i}$}}{top,mid}
\end{fmfgraph*}

%%%%%%%%%%%%%%%%%%%%
%% BUBBLE GRAPHS  

\begin{fmfgraph*}(100,36)
\fmfkeep{bubble}
\fmfleft{in}
\fmfright{out}
\fmfdot{v1}
\fmfdot{v2}
\fmf{dashes_arrow,label=\small{$\omega_\alpha (p)$}}{in,v1}
\fmf{dashes_arrow}{v2,out}
\fmf{plain_arrow,left,tension=0.6,label=\small{$\omega_\alpha (k)$}}{v1,v2}
\fmf{dbl_plain,right,tension=0.6,label=\small{$y(q)$}}{v1,v2}
\end{fmfgraph*}

%\begin{fmfgraph*}(72,25)
%\fmfkeep{tad-nolabel}
%\fmfset{dash_len}{6pt} % this seems to be a local change
%\fmfleft{in,p1}
%\fmfright{out,p2}
%\fmfdot{c}
%\fmf{dbl_dashes,label=\small{$ y_i $}}{in,c}
%\fmf{dbl_dashes}{c,out}
%\fmf{plain_arrow,right, tension=0.8}{c,c}
%\fmf{phantom, tension=0.2}{p1,p2}
%\end{fmfgraph*}

\begin{fmfgraph*}(100,36)
\fmfkeep{bubble-nolabel}
\fmfleft{in}
\fmfright{out}
\fmfdot{v1}
\fmfdot{v2}
\fmf{dashes_arrow,label=\small{$y_i$}}{in,v1}
\fmf{dashes_arrow}{v2,out}
\fmf{plain,left,tension=0.6}{v1,v2}
\fmf{plain,right,tension=0.6}{v1,v2}
%\fmf{plain_arrow,left,tension=0.6}{v1,v2}
%\fmf{plain_arrow,right,tension=0.6}{v1,v2}
\end{fmfgraph*}

\begin{fmfgraph*}(100,36)
\fmfkeep{bubble-y1}
\fmfleft{in}
\fmfright{out}
\fmfdot{v1}
\fmfdot{v2}
\fmf{dbl_dashes,label=\small{$y_1(p)$}}{in,v1}
\fmf{dbl_dashes}{v2,out}
\fmf{plain_arrow,left,tension=0.6,label=\small{$\chi_\pm^{(2)} $}}{v1,v2}
\fmf{plain_arrow,right,tension=0.6,label=\small{$\chi_\pm^{(3)}$}}{v1,v2}
\end{fmfgraph*}

\begin{fmfgraph*}(100,36)
\fmfkeep{bubble-yy}
\fmfset{dash_len}{6pt} % this seems to be a local change
\fmfleft{in}
\fmfright{out}
\fmfdot{v1}
\fmfdot{v2}
\fmf{dashes_arrow,label=\small{$ y(p) $}}{in,v1}
\fmf{dashes_arrow}{v2,out}
\fmf{plain_arrow,left,tension=0.6,label=\small{$\omega_\alpha (k)$}}{v1,v2}
\fmf{plain_arrow,left,tension=0.6,label=\small{$\omega_\alpha (q)$}}{v2,v1}
\end{fmfgraph*}

\begin{fmfgraph*}(100,36)
\fmfkeep{bubble-y2-chi24}
\fmfleft{in}
\fmfright{out}
\fmfdot{v1}
\fmfdot{v2}
\fmf{dashes_arrow,label=\small{$y_2 (p)$}}{in,v1}
\fmf{dashes_arrow}{v2,out}
\fmf{plain_arrow,left,tension=0.6,label=\small{$\chi_\pm^{(2)}$}}{v1,v2}
\fmf{plain_arrow,right,tension=0.6,label=\small{$\chi_\pm^{(4)}$}}{v1,v2}
\end{fmfgraph*}

\begin{fmfgraph*}(100,36)
\fmfkeep{bubble-y2-chi13}
\fmfleft{in}
\fmfright{out}
\fmfdot{v1}
\fmfdot{v2}
\fmf{dashes_arrow,label=\small{$y_2 (p)$}}{in,v1}
\fmf{dashes_arrow}{v2,out}
\fmf{plain_arrow,left,tension=0.6,label=\small{$\chi_\pm^{(1)}$}}{v1,v2}
\fmf{plain_arrow,right,tension=0.6,label=\small{$\chi_\pm^{(3)}$}}{v1,v2}
\end{fmfgraph*}

\begin{fmfgraph*}(100,36)
\fmfkeep{bubble-y2-y24}
\fmfleft{in}
\fmfright{out}
\fmfdot{v1}
\fmfdot{v2}
\fmf{dashes_arrow,label=\small{$y_2 (p)$}}{in,v1}
\fmf{dashes_arrow}{v2,out}
\fmf{plain_arrow,left,tension=0.6,label=\small{$y_2$}}{v1,v2}
\fmf{plain_arrow,right,tension=0.6,label=\small{$y_4$}}{v1,v2}
\end{fmfgraph*}

\begin{fmfgraph*}(100,36)
\fmfkeep{bubble-y3-chi34}
\fmfleft{in}
\fmfright{out}
\fmfdot{v1}
\fmfdot{v2}
\fmf{dashes_arrow,label=\small{$y_3 (p)$}}{in,v1}
\fmf{dashes_arrow}{v2,out}
\fmf{plain_arrow,left,tension=0.6,label=\small{$\chi_\pm^{(3)}$}}{v1,v2}
\fmf{plain_arrow,right,tension=0.6,label=\small{$\chi_\pm^{(4)}$}}{v1,v2}
\end{fmfgraph*}

\begin{fmfgraph*}(100,36)
\fmfkeep{bubble-y3-chi12}
\fmfleft{in}
\fmfright{out}
\fmfdot{v1}
\fmfdot{v2}
\fmf{dashes_arrow,label=\small{$y_3 (p)$}}{in,v1}
\fmf{dashes_arrow}{v2,out}
\fmf{plain_arrow,left,tension=0.6,label=\small{$\chi_\pm^{(1)}$}}{v1,v2}
\fmf{plain_arrow,right,tension=0.6,label=\small{$\chi_\pm^{(2)}$}}{v1,v2}
\end{fmfgraph*}

\begin{fmfgraph*}(100,36)
\fmfkeep{bubble-y3-y34}
\fmfleft{in}
\fmfright{out}
\fmfdot{v1}
\fmfdot{v2}
\fmf{dashes_arrow,label=\small{$y_3 (p)$}}{in,v1}
\fmf{dashes_arrow}{v2,out}
\fmf{plain_arrow,left,tension=0.6,label=\small{$y_3$}}{v1,v2}
\fmf{plain_arrow,right,tension=0.6,label=\small{$y_4$}}{v1,v2}
\end{fmfgraph*}

\begin{fmfgraph*}(100,36)
\fmfkeep{bubble-y4-chi23}
\fmfleft{in}
\fmfright{out}
\fmfdot{v1}
\fmfdot{v2}
\fmf{dashes_arrow,label=\small{$y_4 (p)$}}{in,v1}
\fmf{dashes_arrow}{v2,out}
\fmf{plain_arrow,left,tension=0.6,label=\small{$\chi_\pm^{(i)}$}}{v1,v2}
\fmf{plain_arrow,right,tension=0.6,label=\small{$\chi_\pm^{(i)}$}}{v1,v2}
\end{fmfgraph*}

\begin{fmfgraph*}(100,36)
\fmfkeep{bubble-y4-y23}
\fmfleft{in}
\fmfright{out}
\fmfdot{v1}
\fmfdot{v2}
\fmf{dashes_arrow,label=\small{$y_4 (p)$}}{in,v1}
\fmf{dashes_arrow}{v2,out}
\fmf{plain_arrow,left,tension=0.6,label=\small{$y_i$}}{v1,v2}
\fmf{plain_arrow,right,tension=0.6,label=\small{$y_i$}}{v1,v2}
\end{fmfgraph*}
%%%%%%%%%%%%%%%
%% THE END 

\end{fmffile}
}

\newpage
\setcounter{page}{1}
\tableofcontents{}

\section{Introduction}
Gauge / string dualities offer a fundamentally new view on how to understand strongly coupled systems \cite{Maldacena:1997re,Gubser:1998bc,Witten:1998qj,Aharony:1999ti}.  The most well studied case is the original example of $AdS_5 / CFT_4$ \cite{Maldacena:1997re,Witten:1998qj} which relates string theory on $AdS_5\times S^5$ to super Yang--Mills theory on the four dimensional boundary of $AdS_5$. Another more recent incarnation is $AdS_4 / CFT_3$, this time relating (in a certain limit) type IIA string theory on $AdS_4\times \mathbbm{CP}^3$ to a three-dimensional Chern-Simons matter theory \cite{Aharony:2008ug}. A rather remarkable fact is that most of the mathematical tools developed for $AdS_5 / CFT_4$ turns out to apply almost identically in the more recent $AdS_4 / CFT_3$ duality. As is well known by now, the underlying reason for this similarity of seemingly different theories is the existence of integrable structures. Or in other words, the existence of an infinite set of conserved charges which in principle allows for an exact solution of the spectral problem. The language of integrability allows for a reformulation of the spectral problem in terms of an abstract spin chain. The Hamiltonian acting on the spin chain can be diagonalized using Bethe Ansatz techniques which allows for the spectrum to be written down in a closed form, see \cite{Beisert:2010jr} for a recent review on the subject. In the $AdS_4\times \mathbbm{CP}^3$ case there is one subtlety however which was not present in the $AdS_5\times S^5$ case and which is related to the fact that $AdS_4\times \mathbbm{CP}^3$ is not maximally supersymmetric. The standard proof of integrability of the string worldsheet theory \cite{Bena:2003wd} relies on a supercoset formulation. The supercoset sigma model \cite{Arutyunov:2008if,Stefanski:2008ik} can be obtained from the complete Green-Schwarz superstring \cite{Gomis:2008jt} by (partial) gauge-fixing of the kappa-symmetry. It turns out however, that for certain configurations of the string this gauge-fixing becomes inconsistent \cite{Arutyunov:2008if,Gomis:2008jt} and the supercoset model is not capable of describing all physical fermionic d.o.f. of the string. This is the case for example when the string moves only in the $AdS_4$ subspace or forms an instanton by wrapping $\mathbbm{CP}^1\subset\mathbbm{CP}^3$ \cite{Cagnazzo:2009zh}. This suggests that a more general proof of integrability should be sought which does not rely on the supercoset description. First steps in this direction were taken in \cite{Sorokin:2010wn} where the classical integrability of the full Green-Schwarz string was demonstrated to quadratic order in fermions (the integrability was also shown to higher order in fermions in a truncated model), see also \cite{Cagnazzo:2011at} (and for a slightly different approach see \cite{Uvarov:2012bh}). (A similar problem appears in $AdS_2\times S^2\times T^6$ except there the supercoset model never describes all the physical fermions due to the low amount of supersymmetry, nevertheless the integrability of the Green-Schwarz action has been shown to hold to quadratic order in fermions \cite{Sorokin:2011rr,Cagnazzo:2011at}.)

In \cite{Babichenko:2009dk} an analysis of the integrable structures of yet another duality, namely $AdS_3 / CFT_2$, was initiated. On the string side of the duality we have either $AdS_3\times S^3\times T^4$ or $AdS_3\times S^3\times S^3 \times S^1$ supported by RR-flux. For the first background the dual $CFT_2$ should be a two-dimensional sigma model on a moduli space built out of $Q_1$ instantons in a $U(Q_5)$ gauge theory on $T^4$. This is somewhat natural since $AdS_3\times S^3\times T^4$ arises as the near horizon limit of $Q_1/Q_5$ intersecting $D_1/D_5$ branes, \cite{Elitzur:1998mm,Gauntlett:1998kc,Cowdall:1998bu,Boonstra:1998yu,Papadopoulos:1999tw,Giveon:2003ku,deBoer:1999rh}. On the other hand, the dual theory of $AdS_3\times S^3\times S^3 \times S^1$ remains largely unknown, mainly due to the fact that the supergravity approximation fails to be as useful as in the other examples, see \cite{Gukov:2004ym}. Nevertheless, it is possible to write down a supercoset sigma model for this case whose classical equations of motion allows for a Lax representation which ensures classical integrability \cite{Babichenko:2009dk}. By integrating the Lax connection around a closed loop one gets the monodromy matrix which can be used to generate an infinite tower of conserved charges. The finite gap method can then be used to reformulate the equations of motion in terms of a set of integral equations \cite{Kazakov:2004qf,Beisert:2005bm}. These integral equations in turn can be seen as the semiclassical limit of a set of conjectured quantum Bethe equations which diagonalizes the exact S-matrix on the worldsheet \cite{Staudacher:2004tk,Beisert:2005tm}. While the $AdS_3\times S^3 \times M_4$ solutions of supergravity allow for pure NSNS flux, and thus opens for a exact solution using the representations of chiral algebras \cite{Elitzur:1998mm,Maldacena:2000hw,Maldacena:2000kv,Maldacena:2001km}, the  string appearing in the $AdS_3 / CFT_2$ considered here is supported by RR flux. This means that the proper description is the GS string which is more complicated. What is more, the superisometries of $AdS_5 / CFT_4$ and $AdS_4 / CFT_3$ have 32 and 24 supercharges respectively while the duality at hand only has 16, making it even less symmetric than the higher dimensional examples of integrable gauge string dualities. In this paper we will work with the GS string action up to quadratic order in the fermions. We will see that the subtleties that appear in the $AdS_4\times\mathbbm{CP}^3$ case are also present here. In this case the kappa-symmetry gauge-fixing which reduces the GS string to the supercoset model becomes inconsistent when the string moves only in $AdS_3\subset AdS_3\times S^3\times S^3\times S^1$ or in $AdS_3\times S^3\subset AdS_3\times S^3\times T^4$. However we expect that just as in $AdS_4\times\mathbbm{CP}^3$ and $AdS_2\times S^2\times T^6$ it should be possible to prove the classical integrability of the full GS string to quadratic order in fermions also in this case along the lines of \cite{Sorokin:2010wn,Cagnazzo:2011at}, although we will not address this question in the present paper.

The (super)isometries of $AdS_3\times S^3\times S^3$ form two copies of $D(2,1;\alpha)$ which is an exceptional supergroup with a free parameter $\alpha \in [0,1]$ \cite{Frappat:1996pb}\footnote{The $S^1$ factor is not described by the supergroup and it should be added by hand. One might be tempted to take it as a completely decoupled term in the Lagrangian but this is, however, not the case in the supersymmetric formulation since the fermions couple to all transverse directions through the vielbeins.}. The parameter $\alpha$ enters the invariant bilinear form and can be related to the background geometry though the relation
\bea 
\label{radii} 
\frac{1}{R^2_+}+\frac{1}{R^2_-}=\frac{1}{R^2}
\eea 
where $R_\pm$ are the $S^3$ radii and $R$ is the $AdS_3$ radius. This allows for a trigonometric parameterization as 
\bea 
\label{alpha}
\alpha=\frac{R^2}{R_+^2}=\cos^2\phi,\qquad \frac{R^2}{R_-^2}=\sin^2\phi\,.
\eea
A few special cases are worth mentioning. If we take one of the $S^3$ radii to infinity, we effectively decompactify that part of the geometry. That is, starting from the $AdS_3\times S^3\times S^3\times S^1$ string and sending one of the $R_\pm$ with the $S^1$ radius (which is arbitrary) to infinity we should end up with $AdS_3\times S^3\times T^4$. For this reason it should be possible to write down a sigma model parameterized by $\phi$ (or, equivalently $\alpha$) that can incorporate both backgrounds in one unified description. Indeed, this was done in \cite{Babichenko:2009dk}. What is more, for $\alpha=0,\frac{1}{2},1$ (corresponding to the $T^4$ and equal $S^3$ radii cases), the finite gap method was used to propose a set of quantum Bethe equations \cite{Babichenko:2009dk}. These were subsequently generalized to general $\alpha$ in \cite{OhlssonSax:2011ms}. One motivation for the present paper is to compare and augment the proposals of \cite{Babichenko:2009dk,OhlssonSax:2011ms} with explicit string calculations. While computations have been performed on the string side for the $AdS_3\times S^3\times T^4$ case \cite{David:2008yk,David:2010yg,Lunin:2002fw,Gomis:2002qi,Gava:2002xb,Sommovigo:2003kd} very little has been explicitly computed for general $\alpha$.\footnote{However, see the recent paper \cite{ARXIV:1204.3302} where one-loop effects of spinning and folded string configurations are studied.}

In order to perform any worldsheet calculations, the string Lagrangian is needed and we will use a near BMN expansion up to quartic order in the number of fields (but only quadratic order in fermions). As an independent consistency check we show that the theory is finite at one-loop order in dimensional regularization. As a first test of the conjectured Bethe equations we compare their predictions with energies of string states. While this is only a tree--level computation, it is nevertheless an important consistency check to verify that the spectrum agrees. As we will describe in the paper, we find at least partial agreement. For the rank one sectors the agreement is perfect for arbitrary number of string oscillators (or equivalently, Bethe roots). However, looking at a product $SU(2)\times SU(2)$ sector we find that the string energies in the mixing sector cancel between cubic and quartic contributions from the string Hamiltonian. This implies that in order for the Bethe equations to reproduce the string calculation, the length parameter $L$ of the Bethe equations should not mix the two sectors. This does not necessarily conflict with the results of \cite{Babichenko:2009dk} since these effects would be subleading in $L$. Thus they should not change the semiclassical, $L\rightarrow \infty$, limit and the integral equations of \cite{Babichenko:2009dk} should still be reproduced.

We then set out to investigate some properties of the worldsheet S-matrix. In the string sigma model there are heavy, light and massless modes. While the first two are incorporated by the Bethe equations as fundamental and composite excitations, the massless modes are absent. They do however appear as internal states (as intermediate lines in Feynman diagrams) but it is not possible to assign explicit excitation numbers to them. Thus it might be desirable to extend the Bethe equations in a way so that this is possible. We address this question by showing that the reflection part of the worldsheet S-matrix is zero. This is a nice feature since it, in principle, makes it rather straightforward to add the massless modes by hand as a direct sum. We also collect all the remaining light to light bosonic scattering processes in the appendix.

\subsection*{Outline}

We start out by writing down the Green-Schwarz superstring action to quadratic order in fermions using geometric quantities such as the vielbeins, the spin connection and the RR flux in section \ref{stringsection}. We then discuss the kappa-symmetry gauge-fixing of the action and show that in certain cases the kappa-symmetry fixing which would lead to the supercoset sigma model is not admissible, implying that for certain string configurations the supercoset model is not able to describe all the physical fermionic degrees of freedom on the worldsheet. We use the standard light-cone type kappa-gauge which does not suffer from this problem.

In section \ref{BMNsection} we turn to a perturbative analysis where we start out by fixing the bosonic light-cone gauge \cite{Arutyunov:2005hd,Frolov:2006cc}. We then expand in transverse bosonic and fermionic string coordinates and we write down the theory up to quartic order in fields (but only quadratic order in fermions). As a first consistency check we show that the theory is one-loop finite in dimensional regularization. 

We then turn to an analysis of the Hamiltonian in section \ref{Hsection} by comparing the Bethe equations of \cite{OhlssonSax:2011ms} with string energies. Since the string Hamiltonian has both cubic and quartic interaction terms the actual computation boils down to second order perturbation theory. This, however, can be reformulated in terms of an equivalent first order computation by utilizing a canonical, or unitary, transformation of the Hamiltonian \cite{Frolov:2006cc,Sundin:2008vt}. The classical\footnote{Classical means that we ignore normal ordering effects which, together with terms arising from the unitary transformation, should combine into finite size effects, see \cite{Astolfi:2011bg,Astolfi:2011ju}.} energies we compare with the Bethe equations, comes from string states in an $SU(2)$ and $SU(2)\times SU(2)$ subsector. While we find complete agreement for the rank one sector, we find that there are some issues with the product sector. In order for the Bethe equations to reproduce our findings the length parameter $L$ needs to be different for the two sectors. While we do not fully understand the implications of this, one possible explanation is that we simply have two disconnected spin chains. 

In section \ref{Treesection} we show that the reflection piece of the (bosonic) worldsheet S-matrix is zero. We show this explicitly by computing $2\rightarrow 2$ scatterings (of bosons) on the string worldsheet. While we only present a tree-level computation here, we suspect this to be true in the quantum case also. In the appendix we also compute the scattering and transmission part of the (bosonic) S-matrix. However, since the exact S-matrix is not known we can not compare our findings with anything. 

We end the paper with a short summary and discussion about interesting future problems. 

\section{Green-Schwarz superstring in $AdS_3\times S^3\times S^3\times S^1$}
\label{stringsection}
In this section we write down the Green-Schwarz (GS) superstring action up to quadratic order in fermions. For notational details, see appendix \ref{appendix-parameterization}.

\subsection{GS superstring to quadratic order in fermions in general background}

The action for the GS superstring in a type II supergravity background
(with zero background fermionic fields and NS--NS flux, and constant dilaton $\phi_0$) takes the following form up to quadratic
order in fermions \cite{hep-th/9601109,hep-th/9907202}
\footnote{The string coupling $g$ should be related to the background geometry as $g\sim \sqrt{\lambda}\sim \frac{R^2}{\alpha'}$, where $R$ is the AdS curvature radius. How $\sqrt{\lambda}$ should be defined in the dual $CFT_2$ is not yet known. In most equations we will put $g=1$ for simplicity.}
\begin{equation}
\label{action}
S = -g\int\left(\frac{1}{2}\ast e^Ae_A+i*e^A\,\bar{\Theta}\Gamma_A{\mathcal D}\Theta-ie^A\,\bar{\Theta}\Gamma_A\hat\Gamma{\mathcal D}\Theta\right)\,,
\quad\mbox{where}\quad
\hat\Gamma=\left\{
\begin{array}{c}
\Gamma_{11}\\
\mathbbm1\times\sigma^3
\end{array}
\right.
\begin{array}{c}
(\mathrm{IIA})\\
(\mathrm{IIB})
\end{array}\,.
\end{equation}
The $e^A(X)$ $(A=0,1,\cdots,9)$ are worldsheet pullbacks of the vielbein one-forms of the purely bosonic part of the background ($*$ denotes the worldsheet Hodge-dual), and the generalized covariant derivative acting on the fermions is given by
\begin{equation}
\label{E}
{\mathcal D}\Theta=(\nabla-\frac{1}{8}e^A\,\slashed F\Gamma_A)\ \Theta\quad\mbox{where}\quad \nabla\Theta=(d-\frac{1}{4}\Omega^{AB}\Gamma_{AB})\Theta\,,
\end{equation}
where $\Omega^{AB}$ is the spin connection of the background space-time. The coupling to the RR fields comes through the matrix
\begin{equation}
\label{eq:slashedF}
\slashed F=
e^{\phi_0}\left\{
\begin{array}{c}
-\frac{1}{2}\Gamma^{AB}\Gamma_{11}F_{AB}+\frac{1}{4!}\Gamma^{ABCD}F_{ABCD}\\
{}\\
i\sigma^2\Gamma^AF_A-\frac{1}{3!}\sigma^1\Gamma^{ABC}F_{ABC}+\frac{i}{2\cdot5!}\sigma^2\Gamma^{ABCDE}F_{ABCDE}
\end{array}
\right.
\begin{array}{c}
(\mathrm{IIA})\\
{}\\
(\mathrm{IIB})
\end{array}
\end{equation}
in the type IIA and type IIB cases, respectively. 

The two Majorana-Weyl spinors in the IIA case are described as one 32-component Majorana spinor $\Theta$, and in the IIB case as two 32-component Majorana spinors projected onto one chirality $\Theta^I=\frac{1}{2}(1+\Gamma_{11})\Theta^I$ ($I=1,2$). The Pauli matrices $\sigma^1,\sigma^2$ and $\sigma^3$ act on the IIB SO(2)-indices $I,J=1,2$, which will be suppressed. The Majorana condition implies that the conjugate spinors satisfy
\bea 
\label{spinor-conjugation}
\bar{\Theta}=\Theta^\dagger \Gamma_0=\Theta^t\mathcal C
\eea 
where $\mathcal C$ is the charge conjugation matrix (when needed we use the $\Gamma$-matrix representation given in appendix C of \cite{Babichenko:2009dk}). We now turn to the specific background of interest here, $AdS_3\times S^3\times S^3\times S^1$.

\subsection{GS string in type IIA $AdS_3\times S^3\times S^3\times S^1$}
There are two type II supergravity solutions of the form $AdS_3\times S^3\times S^3\times S^1$ with RR flux. One is in type IIB and has $F_3$ flux while the other is in type IIA and has $F_4$ flux. The type IIB solution arises as the near-horizon geometry of intersecting $D1$ and $D5$-branes. Both solutions preserve 16 supersymmetries and they are easily seen to be related by a T-duality along the $S^1$ direction. Since the fermions in the type IIA case can be grouped into a single 32 component Majorana spinor this case is slightly easier to work with and since both backgrounds describe the same physics we will work with this case.

The $AdS_3\times S^3\times S^3\times S^1$ solution to type IIA supergravity is supported by RR four-form flux of the form
\begin{equation}
F_4 = 2e^{-\phi_0}\left(\frac{1}{3!}e^ce^be^a\varepsilon_{abc}+\cos\phi\ \frac{1}{3!}e^{\hat c}e^{\hat b}e^{\hat a}\varepsilon_{\hat a\hat b\hat c}+\sin\phi\ \frac{1}{3!}e^{c'}e^{b'}e^{a'}\varepsilon_{a'b'c'}\right)e^9\,,
\end{equation}
where $\phi_0$ is the (constant) dilaton and we use units where the $AdS_3$ radius is one. The $D=10$ index $A=0,\ldots,9$ splits up into four sets of indices - an $AdS_3$ index $a=0,1,2$, the first and second $S^3$ indices $\hat a=3,4,5$ and $a'=6,7,8$, and the $S^1$ index $9$ and vielbein $e^9=dy$, where $y$ is the $S^1$ coordinate.

Using the form of $F_4$ in (\ref{eq:slashedF}), we obtain
\begin{equation}
\label{Fslash}
\slashed F=4\gamma_*\Gamma^9(1-\mathcal P)\,,
\end{equation}
where $\gamma_*=\frac{1}{3!}\Gamma^{cba}\varepsilon_{abc}=\Gamma^{012}$, $\gamma_*^2=1$ and $\mathcal P$ is a projection matrix given by
\begin{equation}
\label{calP}
\mathcal P=\frac{1}{2}(1+\cos\phi\ \gamma_*\Gamma^{345}+\sin\phi\ \gamma_*\Gamma^{678})\,.
\end{equation}
This is in fact the projector which singles out the 16 supersymmetries preserved by the background. To see this one can look at the supersymmetry variation of the dilatino. It takes the form
\begin{equation}
\delta\lambda=\Gamma^A\slashed F\Gamma_A\epsilon=8\gamma_*\Gamma^9(1-\mathcal P)\epsilon\,.
\end{equation}
For the 16 supersymmetry parameters which satisfy $\epsilon=\mathcal P\epsilon$ we find $\delta\lambda=0$ which means that these correspond to the supersymmetries preserved by the background. Correspondingly the fermions $\Theta$ can be split into $16+16$ as
\begin{equation}
\vartheta=\mathcal P\Theta\qquad\mbox{and}\qquad\upsilon=(1-\mathcal P)\Theta\,.
\end{equation}
The 16 $\vartheta$ correspond to the supersymmetries preserved by the background and the 16 $\upsilon$ to the broken supersymmetries. We refer to the former as coset fermions and the latter as non-coset fermions since a supercoset formulation only describes the fermions which correspond to unbroken supersymmetries.

Using the explicit form of $\slashed F$ in the action (\ref{action}) we find that the quadratic fermion Lagrangian takes the form (from now on we drop the worldsheet form notation)
\begin{align} 
\label{L2f} \nn
\mathcal L_{2F}&=
i\gamma^{ij}e_i{}^A\,\bar\Theta\Gamma_A\nabla_j\Theta
-i\varepsilon^{ij}e_i{}^A\,\bar\Theta\Gamma_A\Gamma_{11}\nabla_j\Theta
-\frac{i}{2}\gamma^{ij}e_i{}^Ae_j{}^B\,\bar\Theta\Gamma_A\Gamma^{0129}\big(1-\mathcal{P}\big)\Gamma_B\Theta\\ \nn
&\phantom{\quad}
+\frac{i}{2}\varepsilon^{ij}e_i{}^Ae_j{}^B\,\bar\Theta\Gamma_A\Gamma_{11}\Gamma^{0129}\big(1-\mathcal{P}\big)\Gamma_B\Theta\ ,
\end{align}
where $i,j$ are worldsheet indices and $\gamma^{ij}=\sqrt{-g}g^{ij}$ is the Weyl-invariant worldsheet metric satisfying $\det \gamma=-1$.

\subsection{Kappa-symmetry gauge fixing}

The Green-Schwarz superstring action is invariant under local fermionic transformations of the target space coordinates $Z^{\mathcal M}=(X^M,\Theta^\mu)$ which take the form
\begin{eqnarray}
\delta_\kappa Z^{\mathcal M}\,{\mathcal E}_{\mathcal M}{}^\alpha&=&\frac{1}{2}(1+\Gamma)^{  \alpha}_{~ \beta}\,\kappa^{ \beta}(\xi)\,,\qquad {\alpha,\beta}=1,\cdots, 32\ ,
\nonumber\\
\delta_\kappa Z^{\mathcal M}\,{\mathcal E}_{\mathcal M}{}^A&=&0\,,\qquad   A=0,1,\cdots,9\ ,
\end{eqnarray}
where $\kappa^\beta(\xi)$ is an arbitrary 32-component spinor parameter, $(\mathcal E^A,\mathcal E^\alpha)$ are the background supervielbeins and $\frac{1}{2}(1+\Gamma)^\alpha{}_\beta$ is a spinor projection matrix with
\begin{equation}
\Gamma=\frac{1}{2\,\sqrt{-\det{g_{ij}}}}\,\varepsilon^{ij}\,{\mathcal
E}_i{}^A\,{\mathcal E}_j{}^B\,\Gamma_{AB}\,\Gamma_{11}, \qquad
\Gamma^2=1\ ,
\end{equation}
$g_{ij}=\mathcal E_i{}^A\mathcal E_j{}^B\eta_{AB}$ being the induced metric on the worldsheet.

This kappa-symmetry can be used to gauge away 16 of the 32 fermions but exactly which ones may be gauged away may depend on the motion of the string since $\Gamma$ depends on this through the pull-back of the supervielbeins $\mathcal E_i{}^A$. Let us consider a gauge-fixing of the form
\begin{equation}
M\Theta=0\,,
\end{equation}
where $M$ is some $32\times32$ matrix which imposes some $n$-dimensional projection of $\Theta$ to vanish ($n\leq16$). By analyzing a (linearized) kappa-symmetry transformation of this gauge-fixing condition, using the fact that $\mathcal E_\mu{}^\alpha=\delta_\mu^\alpha+\mathcal O(\Theta^2)$, one finds that for the gauge-fixing to be admissible there are essentially two possibilities\footnote{In principle intermediate cases could be considered but they will not be relevant here.}: Either $M$ coincides with the kappa-symmetry projector $\frac{1}{2}(1+\Gamma)$ in an $n$-dimensional subspace of the space it projects on \emph{or} $M$ is independent of $\Gamma$ but\footnote{This is a necessary, but not always sufficient, condition ($M\frac{1}{2}(1+\Gamma)$ still has to have rank $n$).}
\begin{equation}
\mathrm{rank}[M,\Gamma]\geq\frac{n}{2}\,.
\end{equation}
(see also the discussion in section 3 of \cite{Grassi:2009yj}). Let us now consider the implications of this fact for the present case.

If we want to describe the string as a supercoset sigma model we should choose the kappa-symmetry gauge-fixing which sets the 16 non-coset fermions to zero,
\begin{equation}
\upsilon=(1-\mathcal{P})\Theta=0\,.
\end{equation}
According to the above discussion and using the form of $\mathcal P$ in (\ref{calP}) we see that, for generic $\phi$, this gauge choice is not possible if the string moves only in the $AdS_3$ subspace since in that case
\begin{equation}
[\mathcal P,\Gamma]=0\qquad\Rightarrow\qquad\mathrm{rank}[\mathcal P,\Gamma]=0<8\,.
\end{equation}
For the special case $\phi=0$, i.e. $AdS_3\times S^3\times T^4$, the situation is worse and the gauge--fixing is inconsistent if the string motion is in the $AdS_3\times S^3$ subspace. The same holds, of course, for the opposite gauge-fixing which sets the coset fermions to zero, $\vartheta=\mathcal P\Theta=0$. We conclude from this that for these configurations of the string, the physical fermions consist of {\em eight coset} fermions related to supersymmetries and {\em eight non-coset} ones related to broken supersymmetries. Therefore the supercoset sigma model will not describe all physical fermions for these string configurations. This is essentially the same problem that occurs in the case of the $AdS_4\times CP^3$ superstring \cite{Arutyunov:2008if,Gomis:2008jt}.

For this reason we will avoid using the gauge that gives the supercoset model which was used in \cite{Babichenko:2009dk}. We will be interested in the BMN-expansion around a string moving along an $S^1$ in the first $S^3$ factor and an $S^1$ in the second $S^3$ such that they make some angle $\beta$. The case $\beta=0$ corresponds to the string moving only along an $S^1$ in the first $S^3$ while $\beta=\pi/2$ corresponds to the (essentially equivalent) case of the string moving only in the second $S^3$. The kappa-gauge that we will impose is therefore the standard one involving the light-cone $\Gamma$-matrices adapted to the BMN geodesic,
\bea 
\label{kappa-gauge}
\Gamma^+\Theta=0,\qquad \Gamma^\pm=\frac{1}{2}\left(\Gamma^0\pm\big(\cos\beta\,\Gamma^5+\sin\beta\,\Gamma^8\big)\right)\,.
\eea 
The matrix $M$ used in the kappa gauge-fixing can be thought of as the projection matrix $-4\Gamma^-\Gamma^+$ and it is not hard to see that for this string configuration it coincides with the kappa-symmetry projection matrix $\frac{1}{2}(1+\Gamma)$ in the 16-dimensional subspace of positive chirality spinors. Since the chirality projector $\frac{1}{2}(1+\Gamma_{11})$ commutes with the kappa-symmetry projector this gives only 8 instead of the 16 gauge-conditions needed. It therefore appears that this standard gauge-fixing would be incomplete. The resolution of this puzzle is that when we fix also the bosonic light-cone gauge, $x^+\sim\tau$, $x^-$ is fixed by the Virasoro conditions in terms of the other fields and this turns out to remove any would be freedom to perform further kappa-symmetry transformations. Therefore consistency with the Virasoro conditions guarantees that the gauge-fixing is complete also in this case.

\section{Light-cone BMN expansion of the action}
\label{BMNsection}
We will study an expansion in transverse coordinates utilizing a BMN-type expansion \cite{Berenstein:2002jq}. First, we consider the lowest order quadratic theory with $\beta$, the angle the geodesic makes in the $(5,8)$-plane, arbitrary and then when going to higher order in perturbation theory we will consider only the case $\beta=\phi$ for simplicity.

The first thing we should do is fix the residual (bosonic) worldsheet symmetries. The gauge we will employ is a uniform light-cone gauge \cite{Arutyunov:2005hd,Frolov:2006cc}, where the light-cone coordinates adapted to the BMN geodesic are chosen as 
\bea 
\label{light-cone-coordinates}
&& x^\pm=\frac{1}{2}\big(t\pm(\cos\beta\,\varphi_5+\sin\beta\,\varphi_8)\big),\qquad v=\sin\beta\,\varphi_5-\cos\beta\,\varphi_8, \\ \nn 
&& t=x^++x^-,\qquad \varphi_5=\cos\beta(x^+-x^-)+\sin\beta\,v,\qquad \varphi_8=\sin\beta(x^+-x^-)-\cos\beta\,v\ ,
\eea 
with the $\varphi_5$ and $\varphi_8$ being the relevant angle coordinates of $S^3\times S^3$. The light-cone gauge means that we align the worldsheet time coordinate with  $x^+$ through
\bea \label{light-cone-gauge}
x^+=\tau,\qquad p^+=\textrm{ constant }
\eea 
where $p^+$ is the conjugate worldsheet momentum density of $x^-$. In the near BMN limit the gauge-fixed Lagrangian has an expansion in the number of transverse fields as
\bea \label{BMNexpansion}
\mathcal{L}=\mathcal{L}_2+\frac{1}{\sqrt{g}}\mathcal{L}_3+\frac{1}{g}\mathcal{L}_4+...
\eea 
where the subscript denote the number of transverse coordinate in each term. To leading orders in perturbation theory this gauge is also consistent with the conformal gauge, i.e. a flat worldsheet metric. However, at quartic order in the transverse field expansion this gauge fails to hold and we need to add higher order corrections to the worldsheet metric. 

\subsection{Quadratic Lagrangian, $\beta$ arbitrary}
In the BMN limit parameterized by the angle $\beta$ the bosonic terms in the Lagrangian (\ref{action}) reduce, at quadratic order in fields and using conformal gauge $\gamma^{ij}=\eta^{ij}$, to (see Appendix A for the parametrization)
\begin{eqnarray}
\mathcal L_2^B&=&
-\frac{1}{2}
\Big(
\partial_ix_1\partial^ix_1
+\partial_ix_2\partial^ix_2
-x_1^2
-x_2^2
+\partial_ix_3\partial^ix_3
+\partial_ix_4\partial^ix_4
-\cos^2\beta\,\cos^2\phi(x_3^2+x_4^2)
\nonumber\\
&&{}
+\partial_ix_6\partial^ix_6
+\partial_ix_7\partial^ix_7
-\sin^2\beta\,\sin^2\phi(x_6^2+x_7^2)
+\partial_iy\partial^iy
+\partial_iv\partial^iv
\Big)\,.
\end{eqnarray}
The spectrum consists of four pairs of bosons with masses
\begin{equation}
m=\big(1,\,\cos\beta\,\cos\phi,\,\sin\beta\,\sin\phi,\,0\big)\,.
\label{eq:b-spectrum}
\end{equation}
We now turn to the fermionic terms. As can be seen from the Lagrangian (\ref{action}), using (\ref{E}) and (\ref{Fslash}), to leading order the contributing pieces of the vielbein and $\slashed F$ are given by
\bea 
\label{BMNFslash}
e^A\Gamma_A = dx^+\Gamma_+\ ,\qquad \slashed F =2\Gamma^{12+9}\big(1+\cos\beta\cos\phi\Gamma^{1234}+\sin\beta\sin\phi\Gamma^{1267}\big)=
4\Gamma^{12+9}\sum_{ij=\pm}m_{ij}P_{ij}\ ,
\eea 
where $\Gamma_+$ is defined in (\ref{light-cone-gamma}) and
\bea 
P_{\pm \pm}=\frac{1}{4}\big(1\pm\Gamma^{1234}\big)\big(1\pm\Gamma^{1267}\big),\qquad m_{\pm \pm}=\frac{1}{2}\big(1\pm \cos\beta\cos\phi \pm \sin\beta\sin\phi\big)\ .
\eea 
Since these projectors are products of two commuting projectors, which project onto 16-dimensional subspaces, $P_{\pm\pm}$ project onto an 8-dimensional subspace. Fixing the kappa, light-cone and conformal gauge the lowest order Lagrangian for the fermions becomes\footnote{$\Gamma^+\Theta=0$, $x^+=\tau,\gamma^{ij}=\eta^{ij}=(+-)$ and $\varepsilon^{01}=1$.}
\bea 
\label{L2bmn}
\mathcal{L}_2^F=i\bar\Theta \Gamma_+ \partial_0\Theta-i\bar\Theta \Gamma_+\Gamma_{11}\partial_1\Theta
+i\sum_{ij=\pm}m_{ij}\,\bar\Theta\Gamma_+\Gamma^{129}P_{ij}\Theta\ .
\eea 
Thus we see that for generic $\beta$ and $\phi$, there are four two-component fermions $\Theta_{\pm\pm}=P_{\pm\pm}\Theta$ of mass $m_{++},m_{+-},m_{-+}$, and $m_{--}$, respectively. 

For the special case $\beta=\phi$ we see that we have four pairs with masses
\bea 
m=\big(1,\,\cos^2\phi,\,\sin^2\phi,\,0\big)\ ,
\eea 
which coincide with the bosonic mass spectrum of (\ref{eq:b-spectrum}), and therefore the maximum amount of worldsheet supersymmetry is preserved in this case. In order to simplify our analysis we will only consider this case from now on.

\subsection{Quadratic and cubic Lagrangian with $\beta=\phi$}

In order to keep the expressions tractable, we will from now on focus on the case where $\beta=\phi$ \cite{Babichenko:2009dk}. We start out by introducing new variables so that (\ref{L2bmn}) has a nice two-dimensional form. This can be done using the explicit representation of $\Theta$ in appendix \ref{appendix-parameterization}, together with
\begin{align}
&y_1=\frac{1}{\sqrt{2}}(x_1-ix_2),\quad y_2=\frac{1}{\sqrt{2}}(x_3-ix_4),\quad y_3=\frac{1}{\sqrt{2}}(x_6-ix_7),\quad y_4=\frac{1}{\sqrt{2}}(v-ix_9) \\ \nn 
&\chi^\pm_1=\cos\frac{\phi}{2}\,\theta^\pm_1+\sin\frac{\phi}{2}\,\theta^\pm_3,\qquad 
\chi^\pm_2=-\cos\frac{\phi}{2}\,\theta^\pm_2+\sin\frac{\phi}{2}\,\theta^\pm_4, \\ \nn  
&\chi^\pm_3=\sin\frac{\phi}{2}\,\theta^\pm_2+\cos\frac{\phi}{2}\,\theta^\pm_4,\qquad 
\chi^\pm_4=\sin\frac{\phi}{2}\,\theta^\pm_1-\cos\frac{\phi}{2}\,\theta^\pm_3. 
\end{align} 
Rescaling the fermionic fields $\chi_\pm\rightarrow \frac{1}{2\sqrt{2}}\chi_\pm$, and using $\partial_\pm=(\partial_0\pm \partial_1)$, we get
\bea 
\label{bmncoordinateL}
\mathcal{L}_2= i\bar\chi_+^i\partial_- \chi_+^i+i\bar\chi_-^i\partial_+\chi_-^i
+\frac{1}{2}\partial_+ y_i \partial_-\bar{y}_i
+\frac{1}{2}\partial_- y_i \partial_+\bar{y}_i
-m_i^2 y_i\bar{y}_i
-m_i\big(\bar\chi_+^i \chi_-^i+\bar\chi_-^i\chi_+^i\big)
\eea 
where 
\bea &&
m_1=1\ ,\qquad m_2=\cos^2\phi\ ,\qquad 
m_3=\sin^2\phi\ ,\qquad m_4=0\ .
\eea
Thus, all in all, we have $8_B+8_F$ that come in pairs of equal masses.

The conformal and light-cone gauges are also compatible at cubic order and expanding the Lagrangian (\ref{action}) we find ($\slashed F$ is still effectively given by (\ref{BMNFslash}) at this order)
\begin{align}
\label{L3-full1}
\nonumber
\mathcal{L}_3 &= \frac{1}{2\sqrt{2}}\sin 2\phi\Big[-\cos^2\phi\big(\bar\chi_-^4\bar\chi_-^2-\bar\chi_-^1\bar\chi_-^3
+\bar\chi_+^1\bar\chi_+^3-\bar\chi_+^4\bar\chi_+^2\big)y_2 \\
& \phantom{\frac{1}{2\sqrt{2}}\sin 2\phi\Big[\quad}
-i\sin^2\phi\big(\chi_-^3\bar\chi_-^4+\chi_-^2\bar\chi_-^1+\chi_+^3\bar\chi_+^4+\chi_+^2\bar\chi_+^1\big)y_3\\ \nn
& \phantom{\frac{1}{2\sqrt{2}}\sin 2\phi\Big[\quad} -2\big(\chi_-^2\bar\chi_+^3+\chi_+^2\bar\chi_-^3\big)y'_1 + 2\big(\chi_-^2\bar\chi_+^2-\chi_+^3\bar\chi_-^3\big)\dot{y}_4 \\ \nn
& \phantom{\frac{1}{2\sqrt{2}}\sin 2\phi\Big[\quad} + \big(\chi_-^3\bar\chi_+^4-\chi_-^2\bar\chi_+^1\big)(\dot{y}_3+y'_3) + \big(\chi_+^3\bar\chi_-^4-\chi_+^2\bar\chi_-^1\big)(\dot{y}_3-y'_3)\\ \nn 
& \phantom{\frac{1}{2\sqrt{2}}\sin 2\phi\Big[\quad} + i\big(\bar\chi_-^3\bar\chi_+^1+\bar\chi_-^2\bar\chi_+^4\big)(\dot{y}_2+y'_2) + i\big(\bar\chi_-^1\bar\chi_+^3+\bar\chi_-^4\bar\chi_+^2\big)(\dot{y}_2-y'_2)\Big]\\ \nn
& \phantom{\quad} -\frac{1}{\sqrt{2}}\sin 2\phi\ \big(\cos^2\phi\ |y_2|^2-\sin^2\phi\ |y_3|^2\big)\ \dot{y}_4 + \text{h.c.}\ ,
\end{align}
where the hermitian conjugate is defined in the standard way, $\big(\chi_-\bar\chi_+\big)^\dagger = \chi_+\bar\chi_-$. Time and spatial derivatives are denoted by dots and primes, respectively. Also, note that for the $\phi=0$ and $\phi=\pi/2$ cases, the entire cubic Lagrangian vanishes, as expected from the $AdS_3 \times S^3$ coset formulation. 

There are three obvious $U(1)$ charges: one $U(1)_{AdS}$ from the transverse $AdS_3$ and two $U(1)_\pm$ from $S^3\times S^3$. Demanding that the cubic Lagrangian be neutral, we can easily read off the charges of the fields (see table (\ref{tab:charges})).

\begin{table}[h]
\centering
\begin{tabular}{c|c|c|c|c|c|c|c|c}
 & $y_1$ & $y_2$& $ y_3$ & $y_4$ & $\chi_\pm^1$ & $\chi_\pm^2$ & $\chi_\pm^3$& $ \chi_\pm^4$ \\ \hline 
$U(1)_+$ & 0 & 1 & 0 & 0 & 1/2 & 1/2 & 1/2 & 1/2 \\ \hline 
$U(1)_-$ & 0 & 0 & 1 & 0 & 1/2 & -1/2 & -1/2 & 1/2  \\ \hline 
$U(1)_{AdS}$ & 1 & 0 & 0 & 0 & -1/2 & -1/2 & 1/2 & 1/2 
\end{tabular}
\caption{U(1) charges}
\label{tab:charges}
\end{table}

Before we end this section, let us point out one important property of the cubic Lagrangian. It is clear that the only decay processes possible for the heavy modes $y_1$ and $\chi_\pm^{(1)}$ are
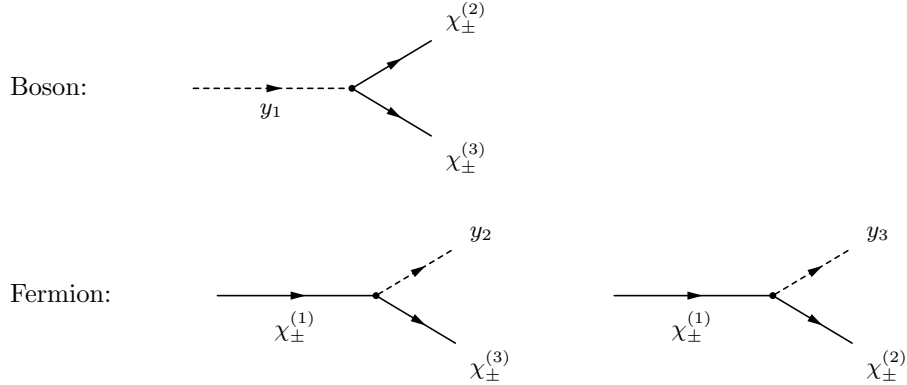
\begin{align*}
&\textrm{Boson:}\qquad \qquad \parbox[top][0.8in][c]{1in}{\fmfreuse{decay1}} & \\ \\ &\textrm{Fermion:}\qquad \qquad \parbox[top][0.8in][c]{1.8in}{\fmfreuse{decay2}}
 \qquad \parbox[top][0.8in][c]{1in}{\fmfreuse{decay3}}
\end{align*}
so the heavy modes decay into two light ones. This property was observed also for the $AdS_4\times \mathbbm{CP}^3$ string which exhibits a composite heavy mode \cite{Zarembo:2009au}. While the above observation of heavy to light-light decay is certainly no proof of a composite heavy mode, it lends support to similar claims made in \cite{Babichenko:2009dk, OhlssonSax:2011ms}. We plan to investigate this question further in an upcoming paper.
%\frac{i}{2\pi}\begin{cases}
%\frac{1}{2\Delta},\qquad\qquad\qquad\qquad\qquad\qquad n=2\quad\mbox{ and } & s=0 \nn \\
%-\frac{1}{\epsilon}+\frac{\gamma_{E}-\log\pi-2\log2}{2}+\frac{1}{2}+\frac{\log\Delta}{2}\;+\mathcal{O}(\epsilon), & s=1\\
%2\Delta\left(-\frac{1}{\epsilon}+\frac{\gamma_{E}-\log\pi-2\log2}{2}+\frac{1}{4}+\frac{\log\Delta}{2}\right)\;+\mathcal{O}(\epsilon), & s=2\,.\end{cases}

\subsection{Gauge-fixing the worldsheet metric}

For technical reasons, it is easiest to fix the light-cone gauge by adding higher-order corrections to the worldsheet metric. As we mentioned, the light-cone gauge
\bea \nn 
x^+=\tau,\qquad p^+=\textrm{constant}
\eea 
is compatible with the conformal gauge for the cubic Lagrangian. However, with the quartic interactions included, the second condition fails to hold. In fact the problem comes only from the purely bosonic part of the Lagrangian. The consistency of the gauge depends on the equation of motion for $x^-$ (which we assume to be at least quadratic in the number of transverse coordinates) and the fermionic contribution comes from 
\bea \nn
-\frac{i}{8}\gamma^{ij}\partial_ix^+\partial_jx^-\left(\bar\Theta\Gamma_+\slashed F \Gamma_-\Theta+\bar\Theta\Gamma_-\slashed F \Gamma_+\Theta\right)+\frac{i}{8}\varepsilon^{ij}\partial_ix^+\partial_jx^-
\left(\bar\Theta\Gamma_+\Gamma_{11}\slashed F \Gamma_-\Theta-\bar\Theta\Gamma_-\Gamma_{11}\slashed F \Gamma_+\Theta\right), 
\eea 
which is obviously zero in the kappa-gauge $\Gamma^+\Theta=0$. Thus, the momentum conjugate to $x^-$ contains no fermionic terms and therefore any modification of the conformal gauge will only contain the bosonic fields.

If we assume that the worldsheet metric receives quadratic corrections as $\gamma=\eta+\hat\gamma$, where $\hat\gamma$ is quadratic in fields, then we find that 
\bea \nn 
\frac{\delta\mathcal{L}}{\delta \dot{x}^-}=-2\hat\gamma^{00}-2|y_1|^2+2\cos^4\phi|y_2|^2+2\sin^4\phi|y_3|^2\ , \qquad 
\frac{\delta\mathcal{L}}{\delta {x}'^-}=-2\hat\gamma^{01}
\eea 
Thus, if we pick\footnote{The second component $\gamma^{11}$ is determined through the condition $\det\gamma=-1$.} 
\begin{align} 
\gamma^{00} &= 1-|y_1|^2+\cos^4\phi|y_2|^2+\sin^4\phi|y_3|^2\ ,\\ \nn
\gamma^{11} &= -1-|y_1|^2+\cos^4\phi|y_2|^2+\sin^4\phi|y_3|^2\ ,
\end{align} 
and $\gamma^{01}=0$, we find $p^+$ constant as required.

Using this, we can immediately write down the full quartic $\mathcal{L}_4$ (except of course for the quartic fermion terms), for arbitrary values of $\phi$. The full Lagrangian is, of course, rather complicated and here we only present its purely bosonic part,
\begin{align}
\label{eq:l4_bosonig}
\nn
\mathcal{L}_4^{B} &= \frac{1}{4}\sin^2 2\phi\left(\cos^2\phi\ |y_2|^2-\sin^2\phi\ |y_3|^2\right)^2 - \frac{1}{8}\sin^2 2\phi\left(\dot{y}_4^2 + \dot{\bar{y}}_4^2 - y_4'^2 - \bar{y}_4'^2\right)\left(|y_2|^2 + |y_3|^2\right)\\ \nn 
&\phantom{\quad} - |\dot{y}_4|^2\left(|y_1|^2 - \cos 2\phi(\cos^2\phi|y_2|^2 - \sin^2\phi|y_3|^2)\right) + |\dot{y}_1|^2\big(\cos^4\phi|y_2|^2 + \sin^4\phi|y_3|^2\big) \\ \nn
&\phantom{\quad} - (|\dot{y}_2|^2 + |\dot{y}_3|^2 + |y'_i|^2)\left(|y_1|^2 - \cos^4\phi|y_2|^2 - \sin^4\phi|y_3|^2\right) - \cos^2\phi|\dot{y}_2|^2|y_2|^2-\sin^2\phi|\dot{y}_3|^2|y_3|^2\\ \nn
&\phantom{\quad} + \cos^2\phi\sin^2\phi|y'_4|^2(|y_2|^2 + |y_3|^2) - |y'_1|^2|y_1|^2 + \cos^2\phi|y'_2|^2|y_2|^2 + \sin^2\phi|y'_3|^2|y_3|^2\,.\\
\end{align} 
In the limiting cases $\phi=0$ and $\phi=\frac{\pi}{2}$, the pure $AdS_3\times S^3$ piece is a direct truncation of the $AdS_5\times S^5$ result given in \cite{Frolov:2006cc}. For the remaining terms relevant for the one-loop computation in the next section, see appendix B.

\subsection{One-loop finiteness}

As a first probe of the quantum consistency of our action for the $AdS_3\times S^3\times S^3\times S^1$ string, we will show that the model is finite in dimensional regularization. That is, if we consider one-loop corrected two-point functions for the bosonic coordinates $y_i$, we observe that all $1/\epsilon$ terms cancel. For arbitrary $\phi$, we have both cubic and quartic interaction terms giving rise to bubble and tadpoles diagrams\footnote{Actually, there are also three vertex tadpoles. For the heavy and massless coordinates these are trivially zero while for the light modes they are zero due to cancellations between boson and fermion loops.}. The tadpoles built out of three-vertices are all zero, and the divergent terms arising from the bubble and tadpole diagrams cancel between each other.

The various loop diagrams encountered are regularized using the standard integral representation
\begin{align} \nn
I_{n}^{s}(\Delta) & = \int d^{d}\ell\,\frac{(\ell^2)^{s}}{\left[\ell^{2}-\Delta\right]^{n}}\\ \nn 
& = i(-1)^s\pi^{d/2}\frac{2}{\Gamma(\frac{d}{2})}\Big(\frac{1}{\Delta}\Big)^{n-s-d/2}\frac{\Gamma(n-s-\frac{d}{2})\Gamma(s+\frac{d}{2})}{2\Gamma(n)}
\end{align}
evaluated at $d=2-\epsilon$. For the bubble diagrams, the divergent integrals are $I^1_2(\Delta)$, corresponding to a logarithmic divergence in hard cutoff. For the tadpoles, on the other hand, we have both $I_1^1(\Delta)$ and $I^0_1(\Delta)$ integrals giving logarithmic and quadratic divergences for a hard cutoff. 

In order to evaluate the contributing diagrams, we sum all the terms arising from the cubic and quartic vertices, where the relevant terms for the latter are collected in (\ref{Lbf}). Since these constitute quite a large number of terms, the actual computation is rather involved, but after some effort we find\footnote{These expressions are evaluated close to the bare pole, $p_0=\sqrt{m_i^2+p_1^2}$.}
\begin{align*}
\mathcal{A}_{B}^i	 & =\parbox[top][0.8in][c]{1.5in}{\fmfreuse{bubble-nolabel}}=\frac{1}{\epsilon}\frac{1}{2\pi}\sin^2 2\phi\,p_1^2+\mathcal{O}(\epsilon^0)
\end{align*}
and
\begin{align*}
\mathcal{A}_{T}^i	 & =\parbox[top][0.8in][c]{1.5in}{\fmfreuse{tad-nolabel}}=-\frac{1}{\epsilon}\frac{1}{2\pi}\sin^2 2\phi\,p_1^2+\mathcal{O}(\epsilon^0)
\end{align*}
where $i$ denotes the bosonic direction. Thus we see that the $1/\epsilon$ terms cancel between the tadpoles and bubbles. In the limiting $\phi=0$ and $\phi=\pi/2$ cases, where there are no cubic terms, we see that the two-point functions are manifestly finite. 

Before ending this section, we should note that in order to determine the finite part of the spectrum, dimensional regularization is not a suitable regulator for the loop integrals \cite{Abbott:2011xp}, see also \cite{Gromov:2008fy,McLoughlin:2008ms,Alday:2008ut,Krishnan:2008zs,McLoughlin:2008he,Bandres:2009kw,Shenderovich:2008bs}. The reason is that in order to maintain unitarity, one should choose a cutoff such that the decay processes, originating from the cubic Lagrangian, are energetically allowed \cite{Astolfi:2011bg} (see also \cite{LopezArcos:2012gb}). The divergent terms, on the other hand, are not sensitive to these issues, but in order to determine the finite part unambiguously, one needs to regularize the theory properly\footnote{There still seems to be a bit of uncertainty in how to regulate the $AdS_4\times\mathbbm{CP}^3$ string properly. In the recent and interesting paper \cite{LopezArcos:2012gb} the authors argue for using a regularization method yielding a different finite result than the unitarity based method.}. We plan to return to this question in future work.

\section{Hamiltonian analysis}
\label{Hsection}

In this section and the next, we will focus on the classical, or tree-level, sector of the string theory. We will start out by calculating energy shifts for an arbitrary length bosonic excitation, and compare this calculation with a conjectured set of Bethe equations. In \cite{Babichenko:2009dk} and \cite{OhlssonSax:2011ms} Bethe equations for $\mathfrak{d}(2,1;\alpha)$ were proposed. These are conjectured to predict the energies of string states for general values of $\phi$. As was the case in $AdS_4\times \mathbbm{CP}^3$, the light modes are the fundamental excitations in the exact solution and the heavier modes are described as composite states of two light modes. How the massless modes enter is not completely clear. For certain simplifying values of $\phi$, the equations seem to capture the full critical spectrum, but in general, the situation seems to require further investigation \cite{OhlssonSax:2011ms}. 

\subsection{Energy shifts}

A very natural set of observables, from a worldsheet point of view, are energy corrections around a BMN vacuum \cite{Berenstein:2002jq}. The way to calculate these for closed strings in various AdS / CFT backgrounds is, by now, a rather well-established procedure \cite{Callan:2004uv, Callan:2004ev, McLoughlin:2004dh,Astolfi:2008ji, Astolfi:2011ju, Hentschel:2007xn, Sundin:2008vt, Sundin:2009zu}. The starting point is the free quadratic BMN Lagrangian, which allows for an exact solution in terms of string oscillators. With the quadratic piece diagonalized, one then calculates the corresponding higher-order corrections to the energy perturbatively.

In order to obtain the energy shifts, we need the string Hamiltonian. We will calculate energy corrections to bosonic modes, mainly focusing on the light modes $y_2$ and $y_3$. This has the pleasant advantage that we only need the purely bosonic Hamiltonian which we can immediately derive from (\ref{bmncoordinateL}), (\ref{L3-full1}) and (\ref{eq:l4_bosonig}) using the Legendre transformation,
\begin{align} 
\label{Hamiltonian}
\mathcal{H} = & |p_i|^2 + |y'_i|^2 + m_i^2|y_i|^2 + \frac{1}{\sqrt{2}}\sin 2\phi\left(p_4 + \bar{p}_4\right)\left(\cos^2\phi\ |y_2|^2-\sin^2\phi\ |y_3|^2\right) \\ \nn 
& + 2|y_1|^2|y'_1|^2
+\cos^2\phi \left(\sin^2\phi\ \left(|p_2|^2+\cos^4\phi\ |y_2|^2\right)-\frac{1}{2}\left(3+\cos 2\phi\right)|y'_2|^2\right)|y_2|^2\\ \nn 
& + \sin^2\phi\left(\cos^2\phi\ \left(|p_3|^2+\sin^4\phi\ |y_3|^2\right)-\frac{1}{2}\left(3-\cos 2\phi\right)|y'_3|^2\right)|y_3|^2 \\ \nn 
& -\cos^4\phi\ |y_2|^2\left(|y'_3|^2 + |p_3|^2\right)-\sin^4\phi\ |y_3|^2\left(|y'_2|^2 + |p_2|^2\right)
-2\cos^4\phi\,\sin^4\phi\ |y_2|^2|y_3|^2\ +...\ ,
\end{align} 
where the ellipses indicate flavor-mixing terms, which will not contribute to our calculation.

The oscillator expansion that diagonalizes the quadratic Hamiltonian is given by
\bea \nn 
y_i = \frac{1}{\sqrt{2\pi}}\int dp\ \frac{1}{2\omega_p^{(i)}}\ \left(a(p)_i\ e^{-ip\cdot \sigma} + b(p)_i^\dagger\ e^{i p\cdot \sigma}\right),\qquad \omega^{(i)}_p = \sqrt{m_i^2 + p^2}.
\eea 
We will calculate the energy corrections to several string states. First, we will consider states built out of one kind of string oscillator 
\bea \label{stringstates1}
\ket{1_A} = \prod_i^A a(p_i)_1^\dagger\ket{0}\ ,\qquad \ket{2_A} = \prod_i^A a(p_i)_2^\dagger\ket{0}\ ,\qquad 
\ket{3_A} = \prod_i^A a(p_i)^\dagger_3\ket{0}\ .
\eea
We shall also consider one more general state which takes values in both $S^3$ spheres. This subsector should constitute a closed $SU(2)\times SU(2)$ sector similar to that of the $AdS_4\times \mathbbm{CP}^3$ string \cite{Minahan:2008hf,Astolfi:2008ji}
\bea 
\label{stringstates2}
\ket{2_A,3_B} = \prod_i^B a(p_i)_{3}^\dagger \prod_j^A a(q_j)_{2}^\dagger\ket{0}\ ,
\eea 
where all mode numbers are distinct for simplicity, and $\ket{0}$ is the BMN vacuum annihilated by all lowering operators. Note that, for both the single flavor and product states, switching oscillators $a_i$ and $b_i$ gives identical results. However, when the fermionic interaction terms are included, some of these states should mix since they are degenerate. The states above will not mix though, since it is not possible to construct other excitations with the same $U(1)$ charges and leading-order energy (see table \ref{tab:charges}). 

Since, for arbitrary $\phi$, we have cubic interactions, we need to consider second-order perturbation theory, either by explicit calculation or by performing a unitary transformation such that the physical information of the cubic piece is rewritten in terms of quartic interactions (see \cite{Frolov:2006cc,Sundin:2008vt}). Evidently, both methods are completely equivalent and importantly, they give rise to terms that need to be regularized. Also, in the case of a nonvanishing cubic piece, the resulting quartic Hamiltonian is most probably $not$ normal ordered. In principle, this gives quadratic normal ordering terms subject to some regularization procedure. The cubic and quartic regularization terms combine into quantum and finite-size effects. In the near-BMN limit, where the coupling is not strictly infinite, the finite-size effects correspond to the finite extension of the string worldsheet. For the $AdS_4\times \mathbbm{CP}^3$ string, these combined into L\"uscher-like finite-size corrections (see \cite{Astolfi:2011ju}). We suspect that the same type of exponentially suppressed terms will appear also for the $AdS_3\times S^3 \times S^3\times S^1$ string. 	

Since almost nothing is known about the quantum theory, we will only consider the classical contribution to the spectrum in this paper. That is, we will simply ignore the terms that need to be regularized (see \cite{Astolfi:2008ji, Sundin:2008vt} for details\footnote{However, note that we expect the spectrum to be exact in the limiting $\phi=0,\pi/2$ cases where the cubic Hamiltonian vanishes.}). Nevertheless, the actual computation is still rather involved. What is more, the unitary transformation we will utilize depends on the massless coordinate, $p_4$. That is, even though the massless terms are not incorporated in the Bethe equations, they still appear as internal lines in Feynman diagrams. Or, as in this case, the massless modes appear as intermediate states in the unitary transformation. Let us explain how the procedure works. The unitary transformation acts on the Hamiltonian as
\bea \nn 
e^{i V}\mathcal{H} e^{-iV}=-\mathcal{H}_3+\textrm{induced quartic terms}
\eea 
and thus, by construction, removes the cubic Hamiltonian at the cost of additional quartic terms. Here we should note a small technical complication. Schematically, the unitary transformation is of the form
\bea \nn
V = g^{-1/2}\sum_{rst}\int dk\, dl\, dm \left[\frac{H_3(k,l,m)^{+++}_{rst}}{\omega^{(r)}(k)+\omega^{(s)}(l)+\omega^{(t)}(m)}
+\frac{H_3(k,l,m)^{++-}_{rst}}{\omega^{(r)}(k)+\omega^{(s)}(l)-\omega^{(t)}(m)} + h.c\right]
\eea 
where the $r,s,t$ sums are over the four bosonic flavors, the $\pm$ labels denote the number of creation/annihilation operators, and the integral is over mode numbers (see \cite{Frolov:2006cc, Sundin:2008vt} for details). Thus, it is clear that for certain values of $k,l,m$, the denominator in the second term can be zero. This is an IR effect and only happens when the mode number of the massless coordinate becomes zero. In order to address this, one should introduce a small non-zero mass, $m_4$, and only in the end send this to zero. 

Using (\ref{Hamiltonian}), together with the method described above, it is straightforward to derive the energy shifts for the states in (\ref{stringstates1}) and (\ref{stringstates2}). A rather lengthy calculation gives (see \cite{Sundin:2008vt} for details)
\begin{align} 
\nn
\Delta E(p_A)_1 &= \frac{1}{4}\sum_{i\neq j}^A \frac{(p_i + p_j)^2}{\omega^{(1)}_i\omega^{(1)}_j}\ ,\\ \nn \\ \nn
\Delta E(p_A)_2 &= - \sum_{i\neq j}^A\left[\frac{\sin^2 2\phi\left(3\cos^4\phi+p_i^2+p_ip_j+p_j^2+\omega^{(2)}_i\omega^{(2)}_j\right)}{16\omega^{(2)}_i\omega^{(2)}_j}\right]_3 
\\ \nn 
& \phantom{- \sum_{i\neq j}^A\ } + \left[\frac{\cos^2\phi\left(\cos^2\phi(p_i+p_j)^2+p_ip_j\sin^2\phi-3\cos^4\phi\sin^2\phi
-\sin^2\phi\,\omega^{(2)}_i\omega^{(2)}_j\right)}{4\omega^{(2)}_i\omega^{(2)}_j}\right]_4\\ \nn 
&= -\frac{\cos^2\phi}{4}\sum_{i\neq j}^A\frac{\left(p_i + p_j\right)^2}{\omega_i^{(2)}\omega_j^{(2)}}\ ,
\end{align} 
\begin{align} 
\nn
\Delta E(p_A)_3 &= -\sum_{i\neq j}^A\left[\frac{\sin^2 2\phi\left(3\sin^4\phi+p_i^2+p_ip_j+p_j^2+\omega^{(3)}_i\omega^{(3)}_j\right)}{16\omega^{(3)}_i\omega^{(3)}_j}\right]_3 
\\ \nn 
& \phantom{- \sum_{i\neq j}^A\ } + \left[\frac{\sin^2\phi\left(\sin^2\phi(p_i+p_j)^2+p_ip_j\cos^2\phi-3\sin^4\phi\cos^2\phi
-\cos^2\phi\,\omega^{(3)}_i\omega^{(3)}_j\right)}{4\omega^{(3)}_i\omega^{(3)}_j}\right]_4\\ \nn 
&= -\frac{\sin^2\phi}{4}\sum_{i\neq j}^B\frac{\left(p_i+p_j\right)^2}{\omega_i^{(3)}\omega_j^{(3)}}\ ,\\ \nn \\ 
\nn
\Delta E(q_A,p_B)_{23} &= -\frac{\cos^2\phi}{4}\sum_{i\neq j}^A\frac{\left(q_i+q_j\right)^2}{\omega_i^{(2)}\omega_j^{(2)}}-\frac{\sin^2\phi}{4}\sum_{i\neq j}^B\frac{\left(p_i+p_j\right)^2}{\omega_i^{(3)}\omega_j^{(3)}} \\ \nn 
&\phantom{\quad}-\frac{1}{2}\sum_i^A\sum_j^B \left\{\frac{-\left[\cos^4\phi\, q_j^2+2\cos^4\phi \sin^4\phi+\sin^4\phi\, p_i^2\right]_3}{\omega^{(2)}_i\omega^{(3)}_j}\right.\\ \nn 
&\left.\phantom{\quad -\frac{1}{2}\sum_i^A\sum_j^B\Big[} +\frac{\left[\cos^4\phi\, q_j^2+2\cos^4\phi \sin^4\phi+\sin^4\phi\, p_i^2\right]_4}{\omega^{(2)}_i\omega^{(3)}_j}\right\} \\  
&=\Delta E(q_A)_2+\Delta E(p_B)_3\ ,
\label{DeltaE}
\end{align} 
where the subscript of the square bracket denotes whether the contribution originates from the cubic or quartic Hamiltonian. While both cubic and quartic contributions are rather involved, it is gratifying to see that the sum of the two simplifies. For the $SU(2)\times SU(2)$ sector, we see that the mixing sector exactly cancels out, and the total energy is just the sum of the two distinct $SU(2)$ sectors\footnote{As can be seen in (\ref{scatterings}), this also happens for S-matrix processes mixing fields from the two $SU(2)$'s.}. Note that for the $y_1$ coordinate, the energy is, up to an overall sign, the same as the $SU(2)$ sector of $AdS_5 \times S^5$ \cite{Frolov:2006cc}. Likewise, for $\phi=0$ and  $\phi=\frac{\pi}{2}$, we see that, up to a sign, the $SL(2)$ result of \cite{Frolov:2006cc} is reproduced.

\subsection{Bethe equations}

The Bethe equations should encode the spectra of both the light and massive coordinates. However, since the heavy mode $y_1$ enters as a composite excitation in the exact solution, it can be rather involved to obtain its solution from the Bethe equations. For this reason, we will only try to reproduce the energies of the light excitations here.

The procedure is as follows: The starting point is the conjectured Bethe equations of \cite{OhlssonSax:2011ms}. These are expressed in terms of Zhukovsky variables $x^\pm$ and the length $L$ of the abstract spin-chain. The ground state of the spin-chain is related to the BMN vacuum which is proportional to $\sqrt{\lambda}>>1$. In order to reproduce the string spectrum one needs to expand the Bethe equations at strong coupling and solve for the rapidity momentum $p_k$ which parameterizes $x^\pm$. Having obtained the (perturbative) solutions for $p_k$ one can then plug this into the magnon dispersion relation which in turn gives a prediction for the energy which we match against the string calculation. For details of this procedure we refer the reader to \cite{Frolov:2006cc,Hentschel:2007xn,Sundin:2008vt}.

\subsection*{SU(2) sector}

The energies (\ref{DeltaE}) of the states (\ref{stringstates1}) should be reproducible from the equations of \cite{OhlssonSax:2011ms} reduced to a rank-one $SU(2)$ sector given by
\bea  \label{BE_rank1}
&& \Big(\frac{x^+_k}{x^-_k}\Big)^L=\prod_{j\neq k}^A \frac{x^+_k-x^-_j}{x^-_k-x^+_j}\times \frac{1-\frac{1}{x^+_kx^-_j}}{1-\frac{1}{x^-_kx^+_j}}\times\sigma^2(x_k,x_j)\,,
\eea
where $L$ denotes the length of the spin-chain. Since we are looking at BMN states, $L\sim g\sim\sqrt{\lambda}$.

While the structural forms of these equations are exactly the same as for the $SU(2)$ spin chain in $AdS_5\times S^5$, the Zhukovsky map is slightly different 
\bea \label{zhukovsky}
x^\pm+\frac{1}{x^\pm}=x+\frac{1}{x}\pm \frac{i\omega_a}{2h(\lambda)}\ ,
\eea 
where
\bea \label{eq:omegas}
\omega_2=2\cos^2\phi,\qquad \omega_3=2\sin^2\phi,
\eea 
depending on the type of excitation. If we use the notation $x^\pm$ and $y^\pm$ to denote excitations with mass $\cos^2\phi$ and $\sin^2\phi$ respectively, then a good parameterization solving (\ref{zhukovsky}) is \cite{Zarembo:2009au}\footnote{A comment on notation: What we call $x^\pm_k$ correspond to $x^\pm_{3,k}$ or $x^\pm_{\bar{3},k}$, while $y^\pm_k$ correspond to $x^\pm_{1,k}$ or $x^\pm_{\bar{1},k}$ in \cite{OhlssonSax:2011ms}.}
\begin{align}\nn 
x^\pm(p_k) &= \frac{\cos^2\phi+\sqrt{\cos^4\phi+4 h(\lambda)^2\sin^2 \frac{p_k}{2}}}{2h(\lambda)\sin\frac{p_k}{2}}\ e^{\pm i\frac{p_k}{2}},\qquad p_k=\frac{p^0_k}{2g}+\frac{p^1_k}{(2g)^2}+...\ ,
\\ \\ \nn
y^\pm(p_k) &= \frac{\sin^2\phi+\sqrt{\sin^4\phi+4 h(\lambda)^2\sin^2 \frac{p_k}{2}}}{2h(\lambda)\sin\frac{p_k}{2}}\ e^{\pm i\frac{p_k}{2}},\qquad q_k=\frac{q^0_k}{2g}+\frac{q^1_k}{(2g)^2}+...
\end{align} 
The function $h(\lambda)$ has a leading order strong-coupling expansion given by \cite{Babichenko:2009dk,OhlssonSax:2011ms}
\bea \nn 
h(\lambda)=\sqrt{\frac{\lambda}{2}}=\frac{g}{2\pi},\qquad \sqrt\lambda,g\,>>\,1\ .
\eea 
For large values of $h(\lambda)$, $\sigma(x_k,x_j)$ is a slightly modified AFS phase \cite{Arutyunov:2004vx, OhlssonSax:2011ms}
\bea \label{AFS}
\sigma(x_l,x_k)=\frac{1-\frac{1}{x^-_lx^+_k}}{1-\frac{1}{x^+_l x^-_k}}\left[\frac{\left(1-\frac{1}{x^+_lx^-_k}\right)\left(1-\frac{1}{x^-_lx^+_k}\right)}{\left(1-\frac{1}{x^+_lx^+_k}\right)\left(1-\frac{1}{x^-_lx^-_k}\right)}
\right]^{\frac{ih}{\omega_a}\left(x_l+\frac{1}{x_l}-x_k-\frac{1}{x_k}\right)}\ .
\eea 
Given a solution of (\ref{BE_rank1}), the corresponding energy and momentum are given by
\bea \label{dilatation}
E = ih(\lambda)\sum_k^A\left(\frac{1}{x^+_k}-\frac{1}{x^-_k}\right),\qquad \prod_k^A \frac{x^+_k}{x^-_k} = 1\ ,
\eea 
where the first equation shows that the magnons have a dispersion relation given by
\bea \label{BE_dispersion}
\epsilon_a = \sqrt{\frac{\omega^2_a}{4} + 4h(\lambda)^2\sin^2\frac{p_i}{2}}\,,
\eea 
where the masses are given in (\ref{eq:omegas}).

In order to solve (\ref{BE_rank1}) we need to express the length, $L$, of the spin chain in terms of string theory variables such as the energy, angular momentum and excitation number ($A$). For $\phi=0$ and $\phi=\frac{\pi}{2}$ the Bethe equations collapse to the rank one equations of $PSU(2,2|4)$ \cite{OhlssonSax:2011ms}. Furthermore, for these two values of $\phi$ the cubic Lagrangian vanishes and the relevant quartic terms are identical to the $AdS_5\times S^5$ case \cite{Frolov:2006cc,Hentschel:2007xn}. Thus, following \cite{Hentschel:2007xn} it becomes clear that the length $L$ is expressed as
\bea\nn
L = g + \frac{1}{2}A-\frac{1}{2}E,\quad \textrm{for}\quad \phi=0,\pi/2\ ,
\eea 
where $E$ is now used to denote the leading-order piece of (\ref{dilatation}), 
\bea 
\nn 
E = \sum_k^A\left(-\frac{\omega}{2} + \sqrt{\frac{\omega^2}{4} + m_k^2}\right) + ... 
\eea 
Focusing on the $\phi=0$ case, we find that the equations collapse to
\bea 
\left(\frac{x^+_k}{x^-_k}\right)^{g+\frac{1}{2}A} = \prod_{j\neq k}^A \frac{x^+_k-x^-_j}{x^-_k-x^+_j} + \mathcal{O}(g^{-2})\ ,
\eea 
and the solution to these equations nicely matches (\ref{DeltaE}) \cite{Frolov:2006cc}. In order to arrive at the above equation, we made use of the following nice identity for the AFS phase \cite{Hentschel:2007xn}:
\bea \label{AFSidentity}
\log\left(\frac{x^+_k}{x^-_k}\right)^{\beta E}\prod_{j\neq k}^A\frac{1-\frac{1}{x^+_kx^-_j}}{1-\frac{1}{x^-_kx^+_j}}\times \sigma^2(x_k,x_j) = \frac{2\pi i}{\omega}\sum_j^A(-1+\beta\omega)\left(-\omega/2+\omega_j\right)p_k + \mathcal{O}(g^{-2})\ ,
\eea 
which vanishes for $\beta=1/2$ and $\phi=0$\footnote{The identity only holds when the momentum constraint (\ref{dilatation}) is satisfied.}. 

In order to reproduce the energies (\ref{DeltaE}) for general $\phi$, the length of the spin chain has to equal
\bea 
L = g + \frac{1}{2}A - \frac{1}{\omega}E\ .
\eea 
We would like to stress that this relation is fixed uniquely, which is easy to see if one, for example, expands in small mode numbers. With this $L$, (\ref{AFSidentity}) is zero and the Bethe equations become
\bea \label{ULCBsu2}
\left(\frac{x^+_k}{x^-_k}\right)^{g+\frac{1}{2}A} = \prod_{j\neq k}^A \frac{x^+_k-x^-_j}{x^-_k-x^+_j}\ ,\qquad 
\left(\frac{y^+_k}{y^-_k}\right)^{g+\frac{1}{2}B} = \prod_{j\neq k}^B \frac{y^+_k-y^-_j}{y^-_k-y^+_j}\ ,
\eea 
and we have the constraints
\bea \nn 
\prod_k\frac{x(p_k)^+}{x(p_k)^-} = 1\qquad \prod_k\frac{y(q_k)^+}{y(q_k)^-} = 1\,.
\eea 
The dispersion relation (\ref{BE_dispersion}) expands as
\bea \label{DispersionExpanded}
E_k^{(i)} = -\frac{\omega}{2}+\omega^{(i)}_k+\Delta E^{(i)}_k\ , \qquad \Delta E(x)^{(i)}_k = \frac{p_k}{8\pi \omega^{(i)}_k}p^1_k,
\eea
where we slightly abused the notation and denote $p_k$ as the mode number of the oscillator state and $p^1_k$ is the subleading piece of the magnon momentum which we solve for using (\ref{ULCBsu2}). Also note that $\omega$ without any subscripts refers to the masses (\ref{eq:omegas}). The index $i$ is either $2$ or $3$ depending on the excitation. Using the explicit solution of $p^1_k$ immediately reproduces the energies of the rank-one sectors, $\Delta E(p_A)_2$ and $\Delta E(p_A)_3$ in (\ref{DeltaE}). 

%Note, if we decompactify one of the $S^3$, that is $\phi=0$ for $\Delta E_2$ and $\phi=\frac{\pi}{2}$ for $\Delta E_3$, then we reproduce the $SL(2)$ spectrum of the $AdS_5\times S^5$ string \cite{Frolov:2006cc} up to a minus sign\footnote{Together with the identifications $P_+=2g$ and $\tilde\lambda=1$.}. This is in nice agreement with \cite{OhlssonSax:2011ms} which showed that the full $d(2,1;\alpha)^2$ equations reduced to the closed $SL(2)$ sector matched the $PSU(2,2|4)$ counterpart up to an overall sign. 

\subsection*{SU(2) $\times$ SU(2) sector}

Here we want to reproduce the $\Delta E_{23}$-shift from the Bethe equations. The largest compact subalgebra of $\mathfrak{d}(2,1,\alpha)$ is $\mathfrak{su}(2) \times \mathfrak{su}(2)$. At weak coupling, the spin chain is that of two decoupled Heisenberg chains related only via the momentum constraint. At strong coupling, we expect the situation to be similar to the $AdS_4\times \mathbbm{CP}^3$ string, which also contains a closed $SU(2)\times SU(2)$ sector \cite{Minahan:2008hf,Astolfi:2008ji}. 

From \cite{Babichenko:2009dk,OhlssonSax:2011ms}, we deduce that the $\Delta E_{23}$-shift should be encoded in
\bea \label{su2su2BE}
&& \left(\frac{x^+_k}{x^-_k}\right)^L = \prod_{j\neq k}^A \frac{x^+_k-x^-_j}{x^-_k-x^+_j}\ \frac{1-\frac{1}{x^+_kx^-_j}}{1-\frac{1}{x^-_kx^+_j}}\ \sigma^2(x_k,x_j)\ ,\\ \nn 
&&
\left(\frac{y^+_k}{y^-_k}\right)^L = \prod_{j\neq k}^B \frac{y^+_k-y^-_j}{y^-_k-y^+_j}\ \frac{1-\frac{1}{y^+_ky^-_j}}{1-\frac{1}{y^-_ky^+_j}}\ \sigma^2(y_k,y_j)\ ,
\eea
augmented with 
\bea 
E = ih\left[\sum_k^A\left(\frac{1}{x^+_k}-\frac{1}{x^-_k}\right) + \sum_k^B\left(\frac{1}{y^+_k}-\frac{1}{y^-_k}\right)\right]\ ,\qquad \prod_k^A\frac{x^+_k}{x^-_k}\prod_k^B\frac{y^+_k}{y^-_k} = 1\ .
\eea 
The parameter $L$ now relates the two equations and following $AdS_4 / CFT_3$ it should be given by \cite{Hentschel:2007xn,Sundin:2008vt,Astolfi:2008ji}
\bea 
L = g+\frac{1}{2}\left(A+B-\frac{1}{\cos^2\phi}\sum^A_kE(x^\pm_k)-\frac{1}{\sin^2\phi}\sum_k^B E(y^\pm_k)\right)\ .
\eea 
If we impose that each subset of mode numbers are separately zero (and distinct), 
\bea \nn 
\prod_k^A\frac{x^+_k}{x^-_k}=\prod_k^B\frac{y^+_k}{y^-_k}=1\ ,
\eea
then (\ref{su2su2BE}) becomes
\begin{align} \label{su2finaleqs} \nn
\left(\frac{x^+_k}{x^-_k}\right)^{g+\frac{1}{2}A} &= \left(\frac{x^+_k}{x^-_k}\right)^{\frac{1}{2}\left(-B+\frac{1}{\sin^2\phi}\sum_j^B E(y^\pm_j)\right)}\prod_{j\neq k}^A \frac{x^+_k-x^-_j}{x^-_k-x^+_j}\ , \\  \\ \nn
\left(\frac{y^+_k}{y^-_k}\right)^{g+\frac{1}{2}B} &= \left(\frac{y^+_k}{y^-_k}\right)^{\frac{1}{2}\left(-A+\frac{1}{\cos^2\phi}\sum_j^A E(x^\pm_j)\right)}\prod_{j\neq k}^B \frac{y^+_k-y^-_j}{y^-_k-y^+_j}\ .
\end{align} 
%Thus, its clear that in order for the Bethe equations to reproduce $\Delta E_{23}$ in (\ref{DeltaE}), then the mixing term has to be given by the first terms on the right hand side of each equation. Parameterizing $x^\pm$ with $q_k$ and $y^\pm$ with $y_k$, gives
%\bea \nn
%&&\log \Big(\frac{x^+_k}{x^-_k}\Big)^{\frac{1}{2}(-B+\frac{1}{\sin^2\phi}\sum_j^B E(y^\pm_j))}=-i\pi q_k\Big[B+\frac{1}{\sin^2\phi}\sum_j^B \big(\sin^2\phi-\omega^{(3)}(p_j)\big)\Big]+... \\ \nn 
%&& \log\Big(\frac{y^+_k}{y^-_k}\Big)^{\frac{1}{2}(-A+\frac{1}{\cos^2\phi}\sum_j^A E(x^\pm_j))}=-i\pi p_k\Big[A+\frac{1}{\cos^2\phi}\sum_j^A \big(\cos^2\phi-\omega^{(2)}(q_j)\big)\Big]+...
%\eea 
Solving the above and using the solutions in (\ref{DispersionExpanded}) we find
\begin{align} 
\label{su2xsu2}
\Delta E &= \Delta E(q_A)_2+\Delta E(p_B)_3\\ \nn 
& \phantom{\quad}-\left[B\sum_k^A\frac{q_k^2}{\omega^{(2)}(q_k)}
+A\sum_k^B\frac{p_k^2}{\omega^{(3)}(p_k)}\right] +\frac{1}{2}\sum_k^A\sum_j^B\frac{\frac{1}{\sin^2\phi}q_k^2\left[\omega^{(3)}(p_j)\right]^2
+\frac{1}{\cos^2\phi}p_j^2\left[\omega^{(2)}(q_k)\right]^2}{\omega^{(2)}(q_k)\omega^{(3)}(p_j)}\ ,
\end{align}
which does not reproduce (\ref{DeltaE}) -- the last line is not zero. Even in the limiting $\phi=\pi/4$ case, we still do not find agreement. We do not know the origin of this mismatch. Perhaps this is a hint that the Bethe equations of \cite{Babichenko:2009dk, OhlssonSax:2011ms} actually describe two spin chains, completely unrelated in the $SU(2)\times SU(2)$ sector. 

We can reconcile the above with the string theory calculation if we assume the parameter $L$ to be distinct in each $SU(2)$. That is, taking 
\bea \nn 
L_2=g+\frac{1}{2}A-\frac{1}{\omega_2} E_2,\qquad L_3=g+\frac{1}{2}B-\frac{1}{\omega_3} E_3
\eea 
for each sector would reproduce the results of (\ref{DeltaE}) since the first terms in the RHS of (\ref{su2finaleqs}) vanish. We would like to stress that the expression for $L$ written above is fairly unique. It is very hard to implement a mixing between the two sectors (for example by adding $B$ and $A$ excitations in $L_2$ and $L_3$ respectively) without contradicting (\ref{DeltaE}) or the S-matrix processes in (\ref{scatterings}). It would be very interesting to investigate this in more detail. For example, one could calculate the full worldsheet S-matrix and from there construct the (string) Bethe equations.

\section{Tree-level scattering}
\label{Treesection}

In order to understand how to properly include the massless modes in the exact solution, we will study how they enter the S-matrix of worldsheet scattering processes. We will study some simple $2\rightarrow 2$ scattering amplitudes for the bosonic particles. Since the exact S-matrix is not known, we are not able to explicitly compare the amplitudes but we do however show that the S-matrix is completely reflectionless. If this is true for the all-loop case, this means the massless modes enter diagonally in the Bethe Ansatz, making it easier to generalize them for the full critical spectrum (see \cite{Beisert:2010jr} and references therein).

Reflectionlessness of the S-matrix is a somewhat unusual property which was also observed for the AdS$_4$ / $CFT_3$ duality \cite{Ahn:2008aa,Ahn:2009zg,Ahn:2009tj}. Under the natural assumption that the S-matrix is also reflectionless at weak coupling, this could shed some light on the unknown CFT$_2$ dual of the $AdS_3 \times S^3 \times S^3\times S^1$ string. 

The worldsheet S-matrix can be separated into three parts: 
\bea \nn 
\textrm{Scattering}\qquad && \mathbbm{S}:\qquad (y y \rightarrow y y)\ , \\ \nn 
\textrm{Transmission}\qquad && \mathbbm{T}:\qquad  (y \bar{y}\rightarrow y \bar{y})\ , \\ \nn 
\textrm{Reflection}\qquad && \mathbbm{R}:\qquad  (y \bar{y}\rightarrow \bar{y} y)\ .
\eea
The S-matrix expands as
\bea \nn 
&& \mathbbm{S}=\mathbbm{1}+iS+...,\qquad 
\mathbbm{T}=\mathbbm{1}+iT+...,\quad 
\mathbbm{R}=iR+...
\eea 
where the contributing diagrams for each part are given by
\begin{align*}
 \mathbbm{S}&=\mathbbm{1}+\frac{i}{g}S+...  \,\,=1+\qquad\parbox[top][0.8in][c]{1in}{\fmfreuse{S-t-chan}} + \qquad \parbox[top][0.8in][c]{1in}{\fmfreuse{S-u-chan}} + \qquad\parbox[top][0.8in][c]{1.5in}{\fmfreuse{S-c}} \\ \nn 
 \mathbbm{T}&=\mathbbm{1}+\frac{i}{g}T+...\,\,  =1+\qquad\parbox[top][0.8in][c]{1in}{\fmfreuse{T-t-chan}} + \qquad \parbox[top][0.8in][c]{1in}{\fmfreuse{T-s-chan}}+ \qquad\parbox[top][0.8in][c]{1.5in}{\fmfreuse{T-c}}
 \\ \nn 
 \mathbbm{R}&=\frac{i}{g}R+... \qquad =\qquad \qquad\parbox[top][0.8in][c]{1in}{\fmfreuse{R-t-chan}} + \qquad \parbox[top][0.8in][c]{1in}{\fmfreuse{R-s-chan}}+ \qquad\parbox[top][0.8in][c]{1.5in}{\fmfreuse{R-c}}
\end{align*}
Below, we will show that the $\mathbbm{R}$ piece is zero for all bosonic $2\rightarrow 2$ scatterings. We provide the light-to-light scattering and transmission components of $\mathbbm{S}$ and $\mathbbm{T}$ in appendix C.

\subsection{Light-to-light reflections}

We start by considering light-to-light processes. In two dimensions, the particles can either keep or exchange their momenta. Except at the special value $\phi=\pi/4$, the masses of $y_2$ and $y_3$ are different which means reflections of these coordinates are trivially zero due to energy conservation,
\bea \nn 
\omega^{(2)}(p_1)+\omega^{(3)}(p_2)\neq \omega^{(2)}(p_2)+\omega^{(3)}(p_1) \qquad \textrm{ when } \qquad \phi\neq \pi/4\ .
\eea 
Thus, the processes we need to consider for general $\phi$ are $y_i \bar{y}_i \rightarrow \bar{y}_i y_i$ and the more general $y_i \bar{y}_j \rightarrow \bar{y}_k y_l$ case at the special point $\phi=\pi/4$.

Ignoring the external leg and overall momentum delta-functions, we find
\bea \nn
\mathbbm{R}\big[y_2\,\bar{y}_2\rightarrow \bar{y}_2 \,y_2\big]:&&\quad 
\Big[4\cos^6\phi\,\sin^2\phi\Big]_c-\frac{1}{2}\sin^22\phi\Big[\cos^4\phi-p_1\,p_2-\omega^{(2)}(p_1)\omega^{(2)}(p_2)\Big]_t \\ \nn 
&&\phantom{\quad}-\frac{1}{2}\sin^22\phi\Big[\cos^4\phi+p_1\,p_2+\omega^{(2)}(p_1)\omega^{(2)}(p_2)\Big]_s = 0\ ,\\ \nn \\ \nn
\mathbbm{R}\big[y_3\,\bar{y}_3\rightarrow \bar{y}_3 \,y_3\big]:&& \quad  
\Big[4\cos^2\phi\,\sin^6\phi\Big]_c-\frac{1}{2}\sin^22\phi\Big[\sin^4\phi-p_1\,p_2-\omega^{(3)}(p_1)\omega^{(3)}(p_2)\Big]_t\\ \nn 
&&\phantom{\quad}-\frac{1}{2}\sin^22\phi\Big[\sin^4\phi+p_1\,p_2+\omega^{(3)}(p_1)\omega^{(3)}(p_2)\Big]_s=0\ ,
\eea 
and for the special case of $\phi=\pi/4$ we have
\bea \nn
\mathbbm{R}\left[y_2\,\bar{y}_3\rightarrow \bar{y}_3 \,y_2\right] = \mathbbm{R}\left[y_3\,\bar{y}_2\rightarrow \bar{y}_2 \,y_3\right]: \qquad\frac{\left[\frac{1}{2}(p_1+p_2)^2\right]_c-\left[\frac{1}{2}(p_1+p_2)^2\right]_t}{1-4p_1\,p_2+4\sqrt{\frac{1}{4}+p_1^2}\sqrt{\frac{1}{4}+p_2^2}}=0\ ,
\eea 
where the subscripts $s$ and $t$ denote the relevant three-vertex diagrams, and $c$ denotes the four-vertex contact contribution. From the above, we thus see that the reflection part of the S-matrix is indeed zero.

\subsection{Light-to-massless reflection}

The presence of the massless modes is a new feature of the AdS$_3$ / CFT$_2$ duality. While they enter as normal excitations on the worldsheet, they are complicated to incorporate in the Bethe Ansatz equations since the finite gap method fails to work. 

In the limiting cases $\phi=0$ and $\phi=\pi/2$, new reflection processes are allowed energetically. For example, at $\phi=0$, the $y_3\,\bar{y}_4\rightarrow\, \bar{y}_4\,y_3$ process is not trivially zero. Of course, the same holds for the other case $\phi=\pi/2$, this time the processes involve $y_2$ and $y_4$. For these special values of $\phi$, the cubic piece (\ref{L3-full1}) vanishes and we only have the contact terms. An easy calculation shows that
\bea \nn 
\mathbbm{R}\left[y_3\,\bar{y}_4\rightarrow \bar{y}_4\,y_3\right]_{\phi=0}=0, \qquad 
\mathbbm{R}\left[y_2\,\bar{y}_4\rightarrow \bar{y}_4\,y_2\right]_{\phi=\pi/2}=0, \\ \nn
\mathbbm{R}\left[y_1\,\bar{y}_2\rightarrow \bar{y}_2\,y_1\right]_{\phi=0}=0, \qquad 
\mathbbm{R}\left[y_1\,\bar{y}_3\rightarrow \bar{y}_3\,y_1\right]_{\phi=\pi/2}=0\ .
\eea 
With this we conclude that the S-matrix of the AdS$_3$ / CFT$_2$ integrable system indeed seems to be reflectionless, at least at tree-level. Of course, to check also the S-matrix for the fermions one would need the action to quartic order in fermions, but supersymmetry suggests that this property should also hold in the fermion sector.

\section{Summary}

We have performed a rather extensive study of the type IIA $AdS_3\times S^3\times S^3\times S^1$ Green-Schwarz superstring up to quadratic order in fermions and discussed issues with fixing its kappa-symmetry. We derived the near BMN expansion of the Lagrangian with quadratic fermions up to quartic order in fields. As a first consistency check, we demonstrated that the one-loop corrections to the two-point functions, built out of the four complex coordinates $y_i$, were finite in dimensional regularization. Both the three- and four-vertex diagrams are separately divergent, but the sum of the two is finite. We then performed a Hamiltonian analysis and compared $SU(2)$ string states with predictions from the conjectured Bethe equations of \cite{OhlssonSax:2011ms}. For the rank-one sectors, we found perfect agreement, while we did not fully understand how to match the product, $SU(2)\times SU(2)$, sector. As it turns out, the string energies arising from the mixing sector exactly cancel between cubic and quartic interaction pieces. This means that, in order for the Bethe equations to reproduce the string calculation, the rank-one equations should decouple completely. A natural way to achieve this is if $L$ has different subleading corrections, distinct for each sector. We are not sure how to interpret this result, and further investigation is probably needed. Note, however, that our result is not necessarily in conflict with \cite{Babichenko:2009dk,OhlssonSax:2011ms} since the subleading effects in $L$ would not show up in the semiclassical limit, and hence the integral equations of \cite{Babichenko:2009dk,OhlssonSax:2011ms} should remain the same.

In the last section of the paper we investigated $2\rightarrow 2$ scattering processes for bosons on the worldsheet. We showed that, at least at tree-level, the two-body S-matrix is reflectionless; this somewhat odd property was also observed in the $AdS_4 / CFT_3$ duality \cite{Ahn:2008aa}. This might be a useful finding if the Bethe equations have to be extended in order to incorporate the massless modes as fundamental excitations.

A natural continuation of the present paper would be to perform a proper quantum computation. While we verified that the theory is one-loop finite, it would definitely be interesting to compute the subleading term in (\ref{BE_dispersion}) from the string theory side. This was, for example, done for the $AdS_4 \times \mathbbm{CP}^3$ string in \cite{Abbott:2011xp}. However, since the worldsheet fields come with different masses, one has to be very careful with the regularization. We plan to return to this question in the future. Another interesting line of research would be to calculate one-loop corrections to the energy along the lines of \cite{Abbott:2010yb}. 

It would also be interesting to verify the integrability of the full GS string (up to quadratic order in fermions) in this background as has been done for $AdS_4 \times \mathbbm{CP}^3$ and $AdS_2\times S^2\times T^6$ \cite{Sorokin:2010wn,Sorokin:2011rr,Cagnazzo:2011at} using similar techniques.

\section*{Acknowledgments}
First and foremost we would like to thank Michael Abbott, Olof Ohlsson Sax, Dima Sorokin and Kostya Zarembo for many interesting discussions, comments and friendly advice. It is also a pleasure to thank Dmitri Bykov, Jeff Murugan and Horatiu Nastase for many stimulating discussions. 

NR is supported by a DAAD scholarship; he wishes to thank PS for giving him the opportunity to contribute to the project. PS acknowledges the financial support of the Claude Leon Foundation. The research of LW was supported in part by NSF grants PHY-0555575 and PHY-0906222.

%-----------------------------APPENDIX-------------------------------

\begin{appendix}
\section*{Appendix}
\section{Notation and parameterization \label{appendix-parameterization}}
The metric on $AdS_3\times S^3\times S^3\times S^1$ is given by
\bea \label{ds2}
ds^2=ds^2(AdS_3)+\frac{1}{\cos^2\phi}ds^2(S^3)+\frac{1}{\sin^2\phi}ds^2(S^3)+ds^2(S^1)
\eea 
where the $\sin$ and $\cos$ factors are there to ensure the triangle identity between the curvature radii (\ref{radii}) and the $AdS_3$ radius is set to one. We choose the following global coordinates \cite{arXiv:0901.4937} 
\bea \label{metric}
ds^2(AdS_3)=-\Big(\frac{1+\frac{1}{4}x_i^2}{1-\frac{1}{4}x_i^2}\Big)^2dt^2+\frac{1}{(1-\frac{1}{4}x_i^2)^2}dx_i^2, \qquad 
ds^2(S^3)=\Big(\frac{1-\frac{1}{4}x_i^2}{1+\frac{1}{4}x_i^2}\Big)^2d\varphi_i^2+\frac{1}{(1+\frac{1}{4}x_i^2)^2}dx_i^2\nn\\
\eea 
where $\varphi_5, \varphi_8$ are the $S^3$ angles which we single out and $\{x_1,x_2\},\{x_3,x_4\},\{x_6,x_7\}, x_9$ are the transverse coordinates. In order to have a smooth interpolation between different values of $\phi$, we will also scale the $S^3$ coordinates as
\bea \nn 
(\varphi_5,x_3,x_4)\rightarrow \cos\phi(\varphi_5,x_3,x_4),\qquad 
(\varphi_8,x_6,x_7)\rightarrow \sin\phi(\varphi_8,x_6,x_7)
\eea 
which allows for nice $T^4$ limits when $\phi=0$ or $\phi=\pi/2$.

The vielbeins can be read off immediately from (\ref{ds2}) and (\ref{metric}). The spin connection of the background is also needed and can be computed from the vanishing of the torsion
\begin{equation}
de^A+e^B\Omega_B{}^A=0\,.
\end{equation}
One finds the non-zero components
\bea \label{spin-connection-components}
&& \Omega^{01}=-\frac{x_1\,dt}{1-\frac{1}{4}x_i^2},\quad \Omega^{02}=-\frac{x_2\,dt}{1-\frac{1}{4}x_i^2},\quad \Omega^{12}=-\frac{1}{2}(x_2e^1-x_1e^2), \\ \nn 
&& \Omega^{35}=-\cos^2\phi\frac{x_3\,d\varphi_5}{1+\frac{\cos^2\phi}{4}x_i^2},
\quad \Omega^{45}=-\cos^2\phi\frac{x_4\,d\varphi_5}{1+\frac{\cos^2\phi}{4}x_i^2},
\quad \Omega^{34}=\cos^2\phi\,\frac{1}{2}(x_4e^3-x_3e^4), \\ \nn 
&& \Omega^{68}=-\sin^2\phi\frac{x_6\,d\varphi_8}{1+\frac{\sin^2\phi}{4}x_i^2},\quad \Omega^{78}=-\sin^2\phi\frac{x_7\,d\varphi_8}{1+\frac{\sin^2\phi}{4}x_i^2},\quad \Omega^{67}=\sin^2\phi\,\frac{1}{2}(x_7e^6-x_6e^7),
\eea 
where $x_i^2$ is $x_1^2+x_2^2$ in the first line, $x_3^2+x_4^2$ in the second line and $x_6^2+x_7^2$ in the third line.

When we work in light-cone coordinates we define
\bea
e^\pm=\frac{1}{2}\big(e^0\pm(\cos\beta\,e^5+\sin\beta\,e^8)\big),\qquad e^v=\sin\beta\,e^5-\cos\beta\,e^8\,,
\eea 
where the angle $\beta$ gives the direction in the $(5,8)$-plane of the geodesic we are interested in.

We will use the $\Gamma$ matrix notation of \cite{Babichenko:2009dk} with the light-cone combinations defined as
\bea \label{light-cone-gamma}
&& \Gamma_\pm=\Gamma_0\pm\big(\cos\beta\,\Gamma_5+\sin\beta\,\Gamma_8\big),\qquad \Gamma_v=\sin\beta\,\Gamma_5-\cos\beta\,\Gamma_8,\qquad \Gamma_{11}=\prod_{i=0}^9\Gamma_i\,.
\eea 
They satisfy
\begin{equation}
\{\Gamma_+,\Gamma_-\}=2\eta_{+-}=-4\,,\qquad \{\Gamma_\pm,\Gamma_v\}=0\,,\qquad\Gamma_v^2=1\,.
\end{equation}
The spinor $\Theta$ satisfying (\ref{spinor-conjugation}) and subject to the gauge fixing condition (\ref{kappa-gauge}) can be decomposed as
\begin{small}
\begin{displaymath}
\Theta=
\left( \begin{array}{c}
-i\sin\beta\,\theta^+_1+i\cos\beta\,\theta^+_3 \\
i\sin\beta\,\theta^+_2+i\cos\beta\,\theta^+_4 \\ 
-\sin\beta\,\bar\theta^+_2-\cos\beta\,\bar\theta^+_4 \\ 
-\sin\beta\,\bar\theta^+_1+\cos\beta\,\bar\theta^+_3 \\ 
\theta^+_3 \\ \theta^+_4 \\ -i \bar\theta^+_4 \\ i \bar\theta^+_3
\end{array} \right) \oplus 
\left( \begin{array}{c}
-i\cos\beta\,\theta^+_1-i\sin\beta\,\theta^+_3 \\
-i\cos\beta\,\theta^+_2+i\sin\beta\,\theta^+_4 \\ 
\cos\beta\,\bar\theta^+_2-\sin\beta\,\bar\theta^+_4 \\ 
-\cos\beta\,\bar\theta^+_1-\sin\beta\,\bar\theta^+_3 \\ 
\theta^+_1 \\ \theta^+_2 \\ -i \bar\theta^+_2 \\ i \bar\theta^+_1
\end{array} \right) \oplus \left( \begin{array}{c}
\theta^-_3 \\ \theta^-_4 \\ i \bar\theta^-_4 \\ -i \bar\theta^-_3 \\ 
i\sin\beta\,\theta^-_1-i\cos\beta\,\theta^-_3 \\
-i\sin\beta\,\theta^-_2-i\cos\beta\,\theta^-_4 \\ 
-\sin\beta\,\bar\theta^-_2-\cos\beta\,\bar\theta^-_4 \\ 
-\sin\beta\,\bar\theta^-_1+\cos\beta\,\bar\theta^-_3 \\ 
\end{array} \right)
\oplus \left( \begin{array}{c}
\theta^-_1 \\ \theta^-_2 \\ i \bar\theta^-_2 \\ -i \bar\theta^-_1 \\ 
i\cos\beta\,\theta^-_1+i\sin\beta\,\theta^-_3 \\
i\cos\beta\,\theta^-_2-i\sin\beta\,\theta^-_4 \\ 
\cos\beta\,\bar\theta^-_2-\sin\beta\,\bar\theta^-_4 \\ 
-\cos\beta\,\bar\theta^-_1-\sin\beta\,\bar\theta^-_3 \\ 
\end{array} \right)
\end{displaymath}
\end{small}

\section{Relevant piece of quartic Lagrangian}

Here we collect the piece of the quartic Lagrangian that is needed for demonstrating one-loop finiteness\footnote{To keep the expression as compact as possible we here denote $\partial_+$ with dot and $\partial_-$ with prime.}
\begin{align}
\label{Lbf}
\mathcal{L}^4_{BF} &= \frac{i}{4}\sum^{4}_{i=1} \Big(\dot{\chi}^i_+\bar{\chi}^i_+ + (\chi^i_-)' \bar{\chi}^i_-\Big)\ |y_1|^2 \\ \nn 
&\quad - \frac{i}{4}\cos^4\phi\ \left[\sum^4_{i=1}\left(\dot{\chi}^i_+\bar{\chi}^i_+ + (\chi^i_-)' \bar{\chi}^i_- \right) - 4i\ \sin^2\phi\ \left(\chi^2_- \bar{\chi}^2_+ - \chi^3_- \bar{\chi}^3_+\right)\right]|y_2|^2\\ \nonumber
&\quad - \frac{i}{4}\sin^4\phi\ \left[\sum^4_{i=1}\left(\dot{\chi}^i_+ \bar{\chi}^i_+ + (\chi^i_-)' \bar{\chi}^i_-\right) + 4i\ \cos^2\phi\ \left(\chi^2_- \bar{\chi}^2_+ - \chi^3_- \bar{\chi}^3_+\right)\right]|y_3|^2\\ \nonumber
&\quad - \frac{1}{2}\left(\chi^1_-\bar{\chi}^1_+ + \cos^2\phi\ \chi^2_-\bar{\chi}^2_+ + \sin^2\phi\ \chi^3_+\bar{\chi}^3_-\right)\dot{\bar{y}}_1 y'_1\\ \nonumber
&\quad - \frac{i}{4}\left[\left(\chi^1_-\bar{\chi}^1_- + \chi^2_-\bar{\chi}^2_- - \chi^3_-\bar{\chi}^3_- - \chi^4_-\bar{\chi}^4_-\right)-\left(\chi^1_+\bar{\chi}^1_+ + \chi^2_+\bar{\chi}^2_+ - \chi^3_+\bar{\chi}^3_+ - \chi^4_+\bar{\chi}^4_+\right)\right]y_1(\dot{\bar{y}}_1 - \bar{y}'_1)\\ \nonumber
&\quad + \frac{1}{2}\left(\cos^2\phi\ \chi^1_+\bar{\chi}^1_- +  \chi^2_+\bar{\chi}^2_- + \sin^2\phi\ \chi^4_+\bar{\chi}^4_-\right)\ \dot{\bar{y}}_2 y'_2 -\frac{i}{4}\cos^2\phi\,\chi^i_-\bar\chi^i_-\,y_2\big(\dot{\bar{y}}_2-\cos^2\phi\bar{y}_2'\big)
\\ \nonumber
&\quad -\frac{i}{4}\cos^2\phi\,\chi^i_+\bar\chi^i_+\,y_2\big(\bar{y}_2'-\cos^2\phi\dot{\bar{y}}_2\big)
+\frac{1}{2}\left(\sin^2\phi\chi^1_+\bar{\chi}^1_- + \chi^3_-\bar{\chi}^3_+ + \cos^2\phi\chi^4_+\bar{\chi}^4_-\right)\dot{\bar{y}}_3 y'_3\\ \nonumber 
&\quad - \frac{i}{4}\sin^2\phi
\left(\chi^1_-\bar{\chi}^1_- - \chi^2_-\bar{\chi}^2_- - \chi^3_-\bar{\chi}^3_- + \chi^4_-\bar{\chi}^4_-\right)\,y_3(\dot{\bar{y}}_3-\sin^2\phi\ \bar{y}'_3)  \\ \nn 
& \quad -\frac{i}{4}\sin^2\phi\left(\chi^1_+\bar{\chi}^1_+ - \chi^2_+\bar{\chi}^2_+ - \chi^3_+\bar{\chi}^3_+ + \chi^4_+\bar{\chi}^4_+\right)\,y_3(\bar{y}'_3-\sin^2\phi\ \dot{\bar{y}}_3)
 \\ \nn 
&\quad + \frac{1}{2}\Big(\sin^2\phi\ \chi^2_-\bar{\chi}^2_+ +\cos^2\phi\ \chi^3_+\bar{\chi}^3_- + \chi^4_-\bar{\chi}^4_+\Big)\dot{y}_4 y'_4 \\ \nn 
&\quad + h.c. + ...\ ,
\end{align}
where the ellipses denote parts not relevant for the computation.

\section{Light to light scattering}
Here we collect the $\mathbbm{S}$ and $\mathbbm{T}$ pieces of the light to light S-matrix, 
\bea \label{scatterings}
&& \textbf{22$\rightarrow$ 22:} \\ \nn 
&& \mathbbm{S}=\frac{1}{2}\sin^2 2\phi\Big[-\cos^4\phi+p_1\,p_2+\omega^{(2)}(p_1)\omega^{(2)}(p_2)\Big]_u 
+\frac{1}{2}\sin^2 2\phi\Big[p_1^2+p_2^2\Big]_t\\ \nn 
&& 
+\cos^2\phi\Big[2\cos^4\phi\sin^2\phi+2\cos^2\phi\big(p_1^2+p_1\,p_2+p_2^2\big)+2p_1\,p_2-2\sin^2\phi\omega^{(2)}(p_1)\omega^{(2)}(p_2)\Big]_c \\ \nn 
&&
=2\cos^2\phi\big(p_1+p_2\big)^2, \\ \nn 
&& \mathbbm{T}=\frac{1}{2}\sin^22\phi\Big[-\cos^4\phi-p_1p_2-\omega^{(2)}(p_1)\omega^{(2)}(p_2)\Big]_s +\frac{1}{2} \sin^2 2\phi\Big[p_1^2+p_2^2\Big]_t\\ \nn 
&& +\cos^2\phi\Big[2\cos^4\phi\sin^2\phi+2\cos^2\phi\big(p_1^2-p_1p_2+p_2^2\big)-2p_1p_2+2\sin^2\phi\omega^{(2)}(p_1)\omega^{(2)}(p_2)\Big]_c
\\ \nn 
&&
=2\cos^2\phi\big(p_1-p_2\big)^2, \\ \nn 
&&\textbf{33$\rightarrow$ 33:} \\ \nn 
&&\mathbbm{S}=\frac{1}{2}\sin^2 2\phi\Big[-\sin^4\phi+p_1\,p_2+\omega^{(3)}(p_1)\omega^{(3)}(p_2)\Big]_u +\frac{1}{2} \sin^2 2\phi\Big[p_1^2+p_2^2\Big]_t\\ \nn 
&& 
+\sin^2\phi\Big[2\sin^4\phi\cos^2\phi+2\sin^2\phi\big(p_1^2+p_1\,p_2+p_2^2\big)+2p_1\,p_2-2\cos^2\phi\omega^{(3)}(p_1)\omega^{(3)}(p_2)\Big]_c \\ \nn
&& =2\sin^2\phi\big(p_1+p_2\big)^2,\\ \nn
&&\mathbbm{T}=\frac{1}{2}\sin^22\phi\Big[-\sin^4\phi-p_1p_2-\omega^{(3)}(p_1)\omega^{(3)}(p_2)\Big]_s +\frac{1}{2} \sin^2 2\phi\Big[p_1^2+p_2^2\Big]_t\\ \nn 
&& +\sin^2\phi\Big[2\sin^4\phi\cos^2\phi+2\sin^2\phi\big(p_1^2-p_1p_2+p_2^2\big)-2p_1p_2+2\cos^2\phi\omega^{(3)}(p_1)\omega^{(3)}(p_2)\Big]_c
\\ \nn 
&&=2\sin^2\phi\big(p_1-p_2\big)^2, \\ \nn 
&& \textbf{23$\rightarrow$ 23:}\qquad \mathbbm{S}=\mathbbm{T}=-2\Big[\cos^4 \phi p_2^2+\sin^4\phi p_1^2\Big]_t+2\Big[\cos^4 \phi p_2^2+\sin^4\phi p_1^2\Big]_c=0,\\ \nn
&&\textbf{32$\rightarrow$ 32:} \qquad
 \mathbbm{S}=\mathbbm{T}=-2\Big[\cos^4 \phi p_1^2+\sin^4\phi p_2^2\Big]_t+2\Big[\cos^4 \phi p_1^2+\sin^4\phi p_2^2\Big]_c=0\,.
\eea 
Note that we have neglected the overall delta functions and external leg factors. As was the case in the Hamiltonian computation, the various contribution tend to cancel among each other. 

Before ending this section we would like to mention that care has to be taken when evaluating the t-channel contributions. Naively one gets $0/0$ expressions and in order to obtain the correct result one should symmetrize over the in and out going momenta and simplify the expressions before enforcing the overall energy and momentum conservation\footnote{We would like to thank Kostya Zarembo for making us aware of these subtleties.}. 
\end{appendix}

\end{document}